\documentclass[a4paper,11pt]{article}
\usepackage{jheppub}
\usepackage{besphysics}
\usepackage{color}
\usepackage{amsmath}
\usepackage{epsfig}
\usepackage{multirow}
\usepackage{overpic}
\usepackage{colortbl}
\usepackage{enumitem}
\newsavebox{\tablebox}
\usepackage{pifont}
\usepackage{enumitem}
\usepackage{tikz}
\usepackage{etoolbox}
\usepackage{comment}
\usepackage[T1]{fontenc}
\usepackage{hyperref}
\usepackage{breakurl}
\usepackage{lineno}
\usepackage[figuresright]{rotating}

\title{\boldmath Measurement of the Electromagnetic Transition Form-factors in the decays $\eta'\rightarrow\pi^+\pi^-l^+l^-$}

\collaboration{The BESIII Collaboration}

\collaborationImg{\includegraphics[height=30mm,angle=90]{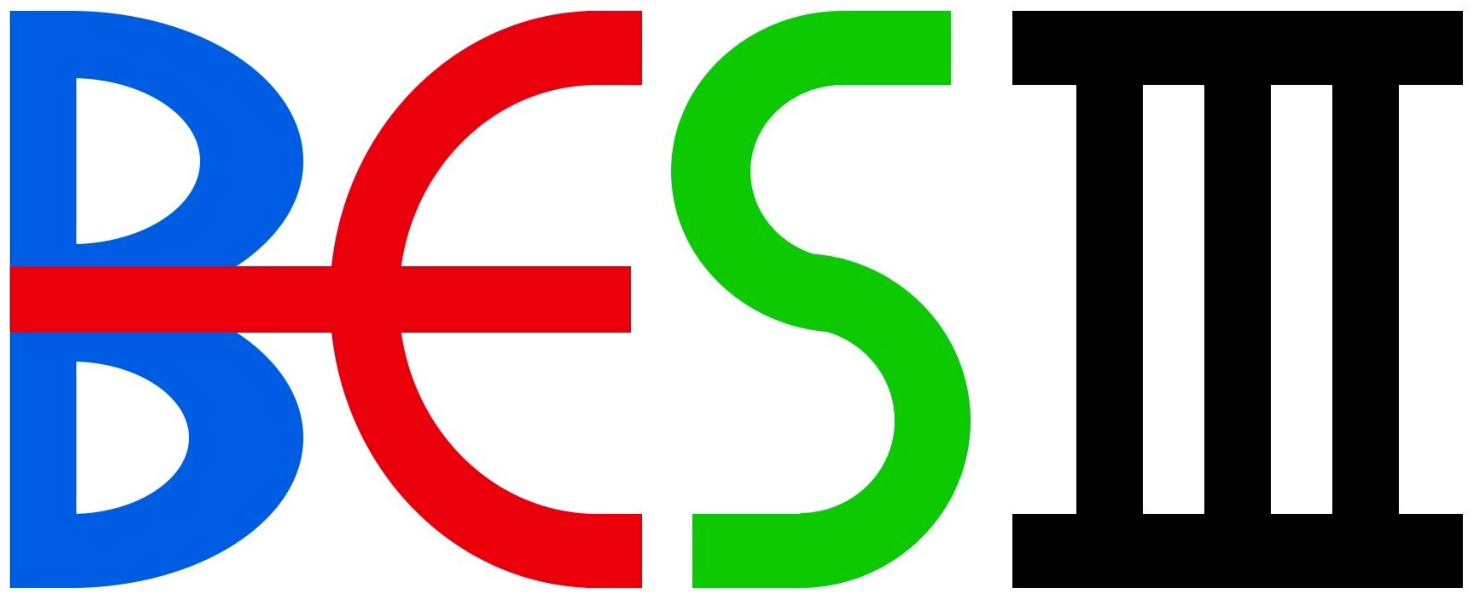}} 

\abstract{
With a sample of $(10087\pm44)\times10^{6}$ $J/\psi$ events accumulated with the BESIII detector, we analyze the decays $\eta'\rightarrow\pi^+\pi^-l^+l^-(l=e,$ $\mu)$ via the process $J/\psi\rightarrow\gamma\eta'$. 
The branching fractions are measured to be $\mathcal{B}(\eta'\rightarrow\pi^+\pi^-e^+e^-)=(2.45\pm0.02(\rm{stat.})\pm0.08(\rm{syst.})) \times10^{-3}$ and $\mathcal{B}(\eta'\rightarrow\pi^+\pi^-\mu^+\mu^-)=(2.16\pm0.12(\rm{stat.})\pm0.06(\rm{syst.}))\times10^{-5}$,
and the ratio is $\frac{\mathcal{B}(\eta'\rightarrow\pi^{+}\pi^{-}e^{+}e^{-})}{\mathcal{B}(\eta'\rightarrow\pi^{+}\pi^{-}\mu^{+}\mu^{-})} = 113.4\pm0.9(\rm{stat.})\pm3.7(\rm{syst.})$.
In addition, by combining the $\eta'\rightarrow\pi^+\pi^-e^+e^-$ and $\eta'\rightarrow\pi^+\pi^-\mu^+\mu^-$ decays, the slope parameter of the electromagnetic transition form factor is measured to be $b_{\eta'}=1.30\pm0.19\ (\mathrm{GeV}/c^{2})^{-2}$, which is consistent with previous measurements from BESIII and theoretical predictions from the VMD model.
The asymmetry in the angle between the $\pi^+\pi^-$ and $l^+l^-$ decay planes, which has the potential to reveal the $CP$-violation originating from an unconventional electric dipole transition, is also investigated.
The asymmetry parameters are determined to be $\mathcal{A}_{CP}(\eta'\rightarrow\pi^+\pi^-e^+e^-)=(-0.21\pm0.73(\rm{stat.})\pm0.01(\rm{syst.}))\%$ and $\mathcal{A}_{CP}(\eta'\rightarrow\pi^+\pi^-\mu^+\mu^-)=(0.62\pm4.71(\rm{stat.})\pm0.08(\rm{syst.}))\%$, implying that no evidence of $CP$-violation is observed at the present statistics.
Finally, an axion-like particle is searched for via the decay $\eta'\rightarrow\pi^+\pi^-a, a\rightarrow e^+e^-$, and upper limits of the branching fractions are presented for the mass assumptions of the axion-like particle in the range of $0-500\ \mathrm{MeV}/c^{2}$.}


\begin{document}
\setcounter{tocdepth}{10}
\maketitle
\let\clearpage\relax
\bibliographystyle{JHEP}
\flushbottom

\section{INTRODUCTION}
The $\eta'\rightarrow\pi^+\pi^-l^+l^-$ (with $l=e$ or $\mu$) decays are of great interest in both theoretical and experimental research as they involve contributions from the box-anomaly of quantum chromodynamics (QCD), and can be used to the transition form factors (TFFs)~\cite{1010_2378}.
The TFF is a momentum transfer $(q^2)$ dependent function, which describes the complex internal structure or the full set of intermediate processes that contribute to a reaction.
At low-momentum transfer, the TFF is particularly important to determine the low-energy parameters of these mesons, such as the slope parameter $b_{\eta'}$~\cite{betap_2022}, which plays a critical role in determining the electromagnetic interaction radius.
The vector meson dominance (VMD) model leads to the expectation $b_{\eta'} \approx \frac{1}{m^2_V}$ for the $P\rightarrow\gamma\gamma^{\star}$ decay~\cite{betap_2022, TFF_theory_1992}.
Also, the TFF is of utmost importance because it determines the size of hadronic quantum corrections in the calculation of the anomalous magnetic moment of the muon, $(g-2)_{\mu}$~\cite{betap_2022}.
Theoretically, the $\eta'\rightarrow\pi^+\pi^-l^+l^-$ decays have been investigated with different models, including the hidden gauge model~\cite{1010_2378}, the chiral unitary approach~\cite{chiral_unitary_approach} and the VMD models~\cite{1010_2378}. 
Experimentally, CLEO and BESIII have measured the branching fractions of these decays using $4.0\times10^{4}$ and $6.8\times10^{6}$ $\eta'$ events, respectively.
Different theoretical predictions and previous experimental results of the branching fractions ($\mathcal{B}$) are listed in Table~\ref{theorybr}.

\begin{table}[htbp]
	\small
	\renewcommand\arraystretch{1.1}
	\begin{center}
		\begin{tabular}{c|c|c}
        \hline
        \hline
        & $\mathcal{B}(\eta'\rightarrow\pi^{+}\pi^{-}e^{+}e^{-})$ $(10^{-3})$ & $\mathcal{B}(\eta'\rightarrow\pi^{+}\pi^{-}\mu^{+}\mu^{-})$ $(10^{-5})$ \\
        \hline
        Hidden gauge~\cite{1010_2378}     & $2.17\pm0.21$              & $2.20\pm0.30$          \\
        VMD~\cite{1010_2378}              & $2.27\pm0.13$              & $2.41\pm0.25$          \\
        Unitary $\chi$PT~\cite{betap_2022} & $2.13^{+0.17}_{-0.31}$      & $1.57^{+0.96}_{-0.75}$ \\
        \hline
        CLEO~\cite{CLEO_2008}             & $2.50^{+1.2}_{-0.9}\pm0.5$ & $<24$             \\
        BESIII~(2013)~\cite{BESIII_2013_etap_pipill}    & $2.11\pm0.12\pm0.15$       & $<2.9$                 \\
        BESIII~(2021)~\cite{BESIII_2020_etap_pipimumu, BESIII_2020_etap_pipiee}    & $2.42\pm0.05\pm0.08$       & $1.97\pm0.33\pm0.19$   \\
        \hline
        \hline
		\end{tabular}
	\end{center}
	\caption{Different theoretical predictions and previous experimental results of $\mathcal{B}(\eta'\rightarrow\pi^{+}\pi^{-}l^{+}l^{-})$.}
	\label{theorybr}
\end{table}	

In 2000, the KTeV Collaboration observed a large $CP$-violation in the distribution of the T-odd angle $\varphi$ in $K^{0}_{L}\rightarrow\pi^{+}\pi^{-}e^{+}e^{-}$ decay~\cite{KL_pipiee_1}, where $\varphi$ is the angle between the $e^{+}e^{-}$ decay plane and the $\pi^{+}\pi^{-}$ decay plane in the $K^{0}_{L}$ center-of-mass system.
The interference between the $CP$-conserving and $CP$-violating amplitudes generates an angular distribution proportional to $\sin{2\varphi}$~\cite{1010_2378}.
The asymmetry was first explored in the decay $\eta'\rightarrow\pi^{+}\pi^{-}e^+e^-$ in the BESIII experiment, and the $CP$-violating asymmetry ($\mathcal{A}_{CP}$) was determined to be $(2.9\pm3.7\pm1.1)\%$~\cite{BESIII_2020_etap_pipiee}, which is consistent with zero. 

Additionally, the hadronic decay channels of $\eta'$ mesons could include a signal of a QCD axion, dark photon or other hadronically coupled Axion-like particle (ALP).
Among the most promising modes are the three-body final states $\pi^+\pi^-a$, which have been studied recently~\cite{axion_etaTopipia_2020}.
The ATOMKI experiment reported an anomaly in the angular correlation spectra in $^{8}$Be~\cite{axion_Be8_2016} and $^{4}$He~\cite{axion_He4_2019} nuclear transitions.
A light pseudoscalar particle $a$ decaying to $e^+e^-$~\cite{BUTTAZZO2021136310, Ellwanger:2016wfe, Liu:2021wap} was proposed to explain the anomaly. 
An ALP could also cause a deviation from the expected value of the electron anomalous magnetic moment~\cite{doi:10.1126/science.aap7706, Morel:2020dww, NA64:2021xzo}.

With ten billion $J/\psi$ data events collected with the BESIII detector during 2009-2019~\cite{BESIII_2021_njpsi}, we can improve the precision of the branching fraction measurements, measure the TFFs and search for $CP$-violation and hadronically coupled ALPs.

\section{BESIII DETECTOR}
The BESIII detector~\cite{BESIII_2009_detector} records symmetric $e^+e^-$ collisions 
provided by the BEPCII storage ring~\cite{BESIII_2016_detector}
in the center-of-mass energy range from 2.0 to 4.95~GeV,
with a peak luminosity of $1 \times 10^{33}\;\text{cm}^{-2}\text{s}^{-1}$ 
achieved at $\sqrt{s} = 3.77\;\text{GeV}$. 
BESIII has collected large data samples in this energy region~\cite{BESIII:2020nme}. The cylindrical core of the BESIII detector covers 93\% of the full solid angle and consists of a helium-based multilayer drift chamber~(MDC), a plastic scintillator time-of-flight
system~(TOF), and a CsI(Tl) electromagnetic calorimeter~(EMC),
which are all enclosed in a superconducting solenoidal magnet
providing a 1.0 T magnetic field.
The magnetic field was 0.9~T in 2012, which affects 11\% of the total $J/\psi$ data.
The solenoid is supported by an
octagonal flux-return yoke with resistive plate counter muon
identification modules interleaved with steel. 
The charged-particle momentum resolution at $1~{\rm GeV}/c$ is
$0.5\%$, and the 
${\rm d}E/{\rm d}x$
resolution is $6\%$ for electrons
from Bhabha scattering. The EMC measures photon energies with a
resolution of $2.5\%$ ($5\%$) at $1$~GeV in the barrel (end cap)
region. The time resolution in the TOF barrel region is 68~ps, while
that in the end cap region was 110~ps. 
The end cap TOF
system was upgraded in 2015 using multigap resistive plate chamber
technology, providing a time resolution of
60~ps~\cite{BESIII_2019_detector},
which benefits 87\% of the data used in this analysis.

\section{DATA SAMPLE AND MONTE CARLO SIMULATION}
This analysis is based on $(10087\pm44)\times10^{6}$ $J/\psi$ events collected with the BESIII detector~\cite{BESIII_2021_njpsi}, which yields a sample of about $5.3\times10^{7}$ $\eta'$ events though the radiative decay $J/\psi\rightarrow\gamma\eta'$.

The estimation of background contributions and the determination of detection efficiencies are performed with Monte Carlo (MC) simulations. The BESIII detector is modeled with GEANT4~\cite{GEANT4_2002_1, GEANT4_2006_2, Huang:2022wuo}. The production of the $J/\psi$ resonance is implemented with MC event generator KKMC~\cite{kkmc_1999_1, kkmc_2000_2}, while the decays are simulated by EVTGEN~\cite{EVTGEN_2008, Lange:2001uf}. The possible backgrounds are studied using a sample of $J/\psi$ inclusive events in which the known decays of $J/\psi$ are modeled with branching fractions being set to the world average values given by the Particle Data Group (PDG)~\cite{PDG_2022}, while unknown decays are generated with the LUNDCHARM model~\cite{LUNDCHARM_2000, Yang:2014vra}. 
Specific generators have been developed in this analysis, and the corresponding theoretical models are listed in Table~\ref{Generators}.

\begin{table}[htbp]
	\small
	\renewcommand\arraystretch{1.1}
	\begin{center}
		\begin{tabular}{c|c}
		\hline
		\hline
        Decay mode & Generator model \\
        \hline
        $J/\psi\rightarrow\gamma\eta'$                      & Helicity amplitude\\
        $\eta'\rightarrow\pi^{+}\pi^{-}l^+l^-$              & VMD model~\cite{Event_generators_2012_etap_pipill}\\
        $\eta'\rightarrow\pi^{+}\pi^{-}\pi^{+}\pi^{-}$      & ChPT and VMD model~\cite{Event_generators_2014_etap_4pi}\\
        $\eta^{(\prime)}\rightarrow\gamma\pi^{+}\pi^{-}$    & Amplitude analyses~\cite{BESIII_etap_gpipi}\\
        $\eta'\rightarrow\eta\pi^{+}\pi^{-}$                & Dalitz plot analyses~\cite{Event_generators_2017_etap_pipieta}\\
        $\eta\rightarrow\gamma\mu^+\mu^-$                   & Electromagnetic form factor~\cite{Event_generators_2015_etap_gmumu}\\
		\hline
		\hline
		\end{tabular}
	\end{center}
	\caption{Generator models used for MC simulations.}
	\label{Generators}
\end{table}	

\section{EVENT SELECTION AND BACKGROUND ANALYSIS}
The final state of interest is studied through the decay chain $J/\psi\rightarrow\gamma\eta', \eta^\prime\rightarrow\pi^{+}\pi^{-}l^{+}l^{-}$. Each event is required to contain at least one good photon candidate, and four charged track candidates with a total charge of zero. 
Charged tracks detected in the MDC are required to be within a polar angle $(\theta)$ range of $|\cos \theta| \leq 0.93$, where $\theta$ is defined with respect to the $z$-axis, which is the symmetry axis of the MDC.
Each charged track is required to have the point of closest approach to the interaction point (IP) within $\pm 1$ cm in the plane perpendicular to beam direction and within $\pm 10$ cm in $z$ direction.  
 
Photons are reconstructed from showers in the EMC exceeding a deposited energy of $25\ \mathrm{MeV}$ in the barrel region ($|\cos \theta| < 0.8$) and 50 MeV in the endcap regions $(0.86 < |\cos \theta|< 0.92)$. The angle between the shower position and any charged tracks extrapolated to the EMC must be greater than 15 degrees. Finally, photons are required to arrive within 700 ns from the event start time in order to reduce background from photons that do not originate from the same event.

For each signal candidate, particle identification (PID) is performed using the TOF and d$E$/d$x$ information, and four-constraint (4C) kinematic fits are performed imposing energy and momentum conservation under the hypotheses of $\gamma\pi^{+}\pi^{-}e^{+}e^{-}$ and $\gamma\pi^{+}\pi^{-}\mu^{+}\mu^{-}$ final states.
The variable, $\chi^2_{\rm{4C+ PID}}=\chi^2_{\rm{4C}}+\sum_{i=1}^4\chi^2_{{\rm PID}(i)}$, is used to select the best candidate if the event contains more than one. 
Here $\chi^2_{\rm{4C}}$ and $\chi^2_{\rm{PID}}$ represent the Chi-square of the 4C kinematic fit and PID, respectively, and $i$ corresponds to the charged track candidates in each hypothesis (pion, electron, or muon). 
For each event, the candidate with the smallest $\chi^2_{\rm{4C+ PID}}(\pi^{+}\pi^{-}l^{+}l^{-})$ is kept for further analysis. 

\begin{figure}[htbp]
	\centering
	\begin{overpic}[width=0.49\textwidth]{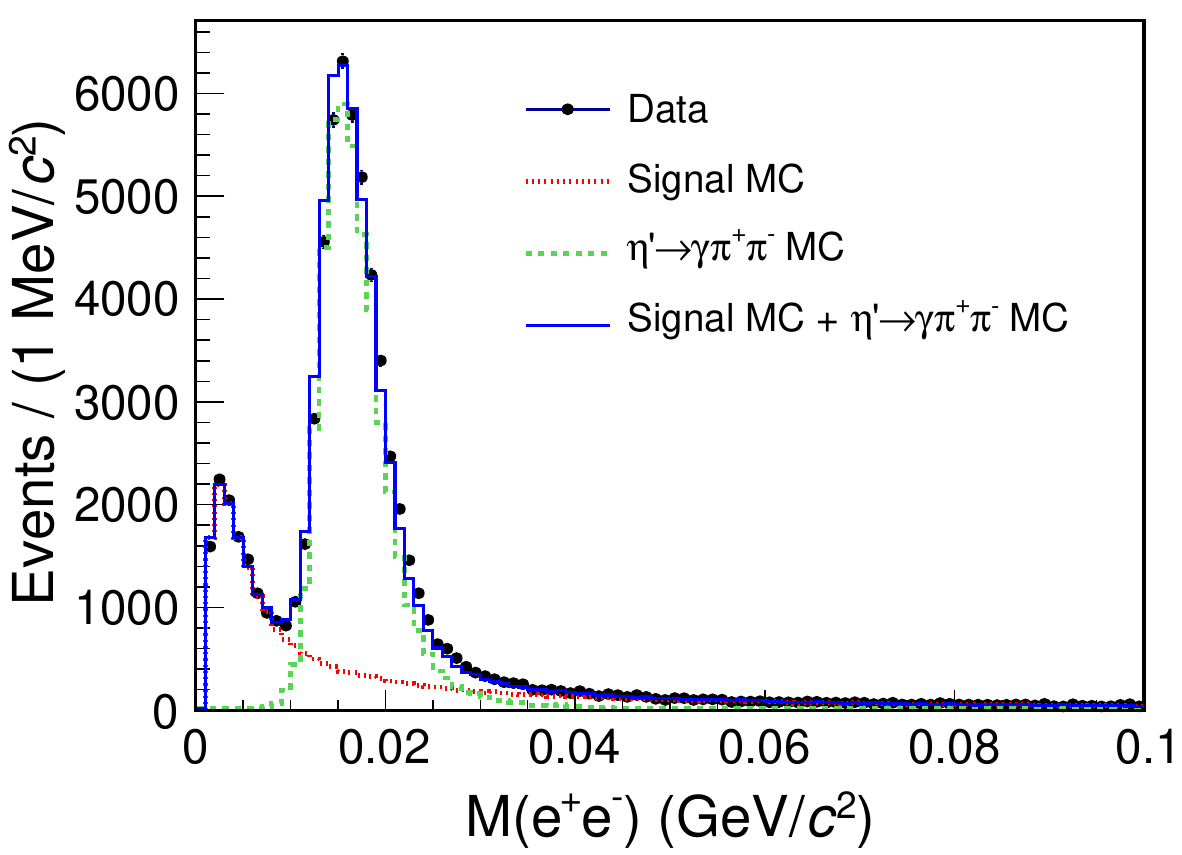}
		\put(80,60){$\bf (a)$}
	\end{overpic}
	\begin{overpic}[width=0.49\textwidth]{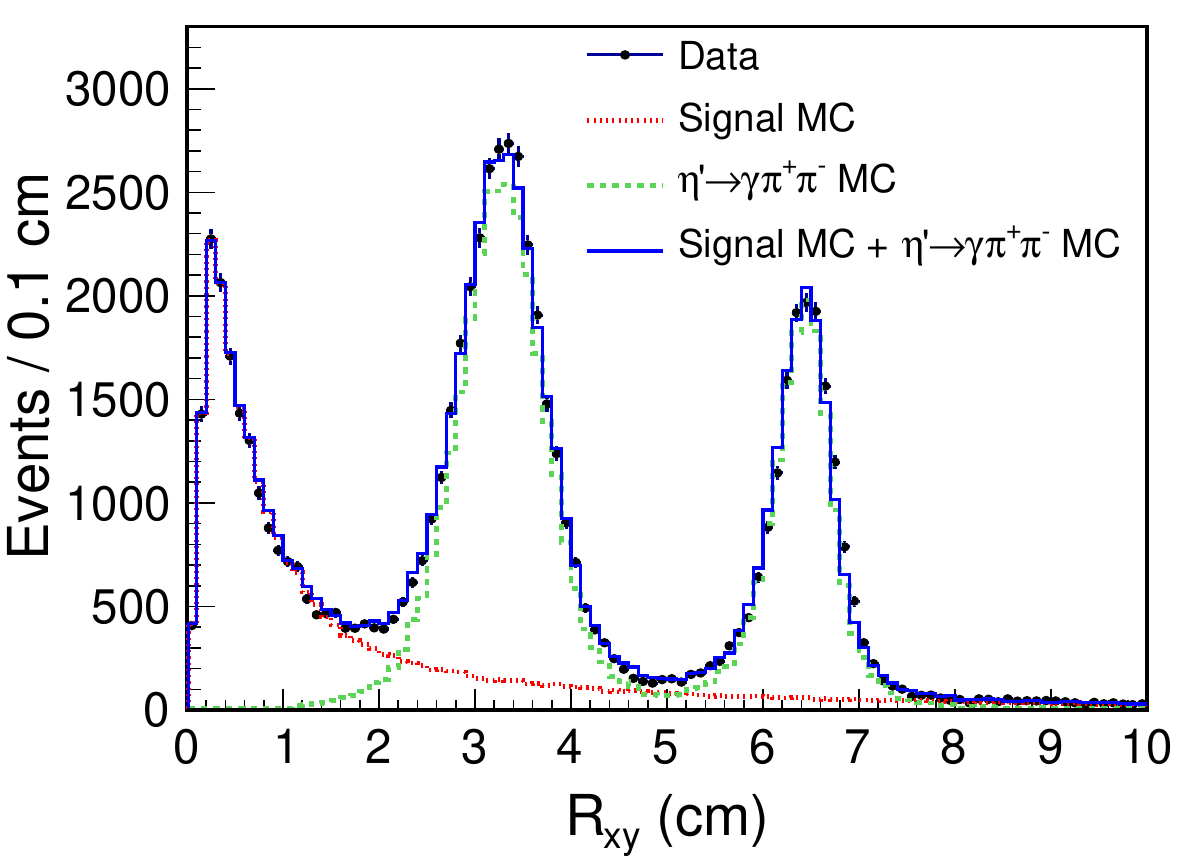}
		\put(25,60){$\bf (b)$}
	\end{overpic}
	\caption{The distributions of (a) the $e^{+}e^{-}$ invariant mass spectrum and (b) the distance of the $e^{+}e^{-}-$vertex from the IP in the transverse direction.
    The dots with error bars represent the data, the red dotted histograms are the MC signal shapes, the green dashed histograms are the $J/\psi\rightarrow\gamma\eta', \eta'\rightarrow\gamma\pi^+\pi^-$ MC shapes, and the blue solid histograms are the sums of the MC signal and MC background from $J/\psi\rightarrow\gamma\eta', \eta'\rightarrow\gamma\pi^+\pi^-$. 
    Both of these MC simulations are normalized to the yields found in Table~\ref{backgrounds}.}	
	\label{Mee_Rxy}
\end{figure}

For the $\eta'\rightarrow\pi^{+}\pi^{-}e^{+}e^{-}$ decay, $\chi^{2}_{\rm{4C+ PID}}(\gamma\pi^+\pi^-e^+e^-)<60$ is required, which has been optimized with respect to the figure of merit $N_{S}/\sqrt{N_{D}}$, where $N_{S}$ is the number of events in the signal MC sample, and $N_{D}$ is the number of events in data.
The primary peaking background comes from $J/\psi\rightarrow\gamma\eta', \eta'\rightarrow\gamma\pi^+\pi^-$ events,
where a photon converts to an $e^{+}e^{-}$ pair at the beam pipe or the inner wall of the MDC.
One might expect the invariant mass of such a conversion pair, $M(e^+e^-)$, to be close to zero for this kind of background.
However, the BESIII tracking algorithm uses the IP as a reference point for all tracks, which means that the direction of tracks that originate elsewhere are mis-reconstructed.
Hence, a conversion pair gains an artificial opening angle, and the reconstructed invariant mass is larger than the true value.
Therefore, the conversion background appears as a large peak at about 0.015 GeV/$c^{2}$ in the $M(e^+e^-)$ distribution, shown in Fig.~\ref{Mee_Rxy} (a). 

The intersection of the $e^+$- and $e^-$-helices in the $r-\phi$ projection provides the distance from the $e^+e^-$-vertex position to the IP, $R_{xy}$~\cite{GamConv}.
Figure~\ref{Mee_Rxy} (b) shows the $R_{xy}$ distribution for the selected events, where $e^+e^-$ pairs peak at three characteristic locations in the detector. 
The signal pairs originate from the IP, whereas the conversion background pairs are at $R_{xy}\approx3$ cm, corresponding to the beam pipe, and at $R_{xy}\approx6$ cm, corresponding to the inner wall of the MDC.

\begin{figure}[htbp]
	\centering
	\begin{overpic}[width=0.49\textwidth]{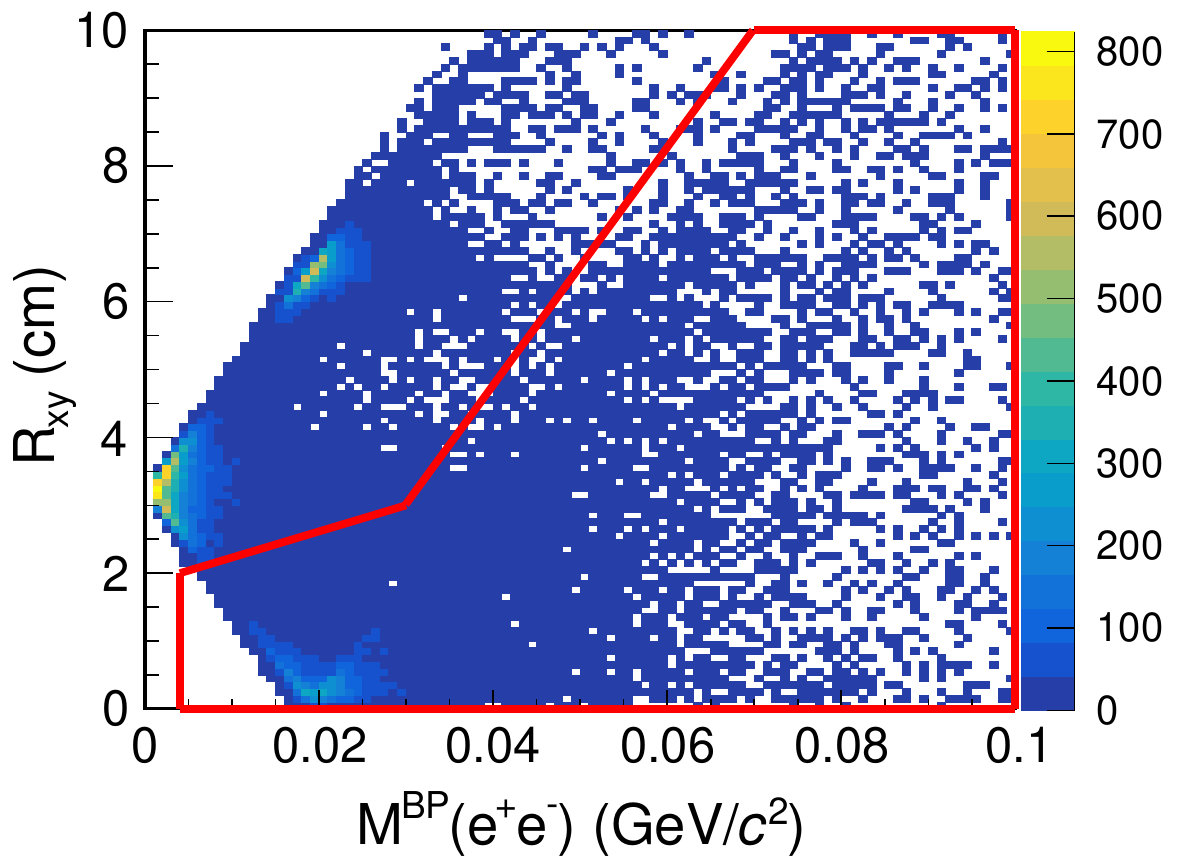}
		\put(20,60){$\bf (a)$}
	\end{overpic}
	\begin{overpic}[width=0.49\textwidth]{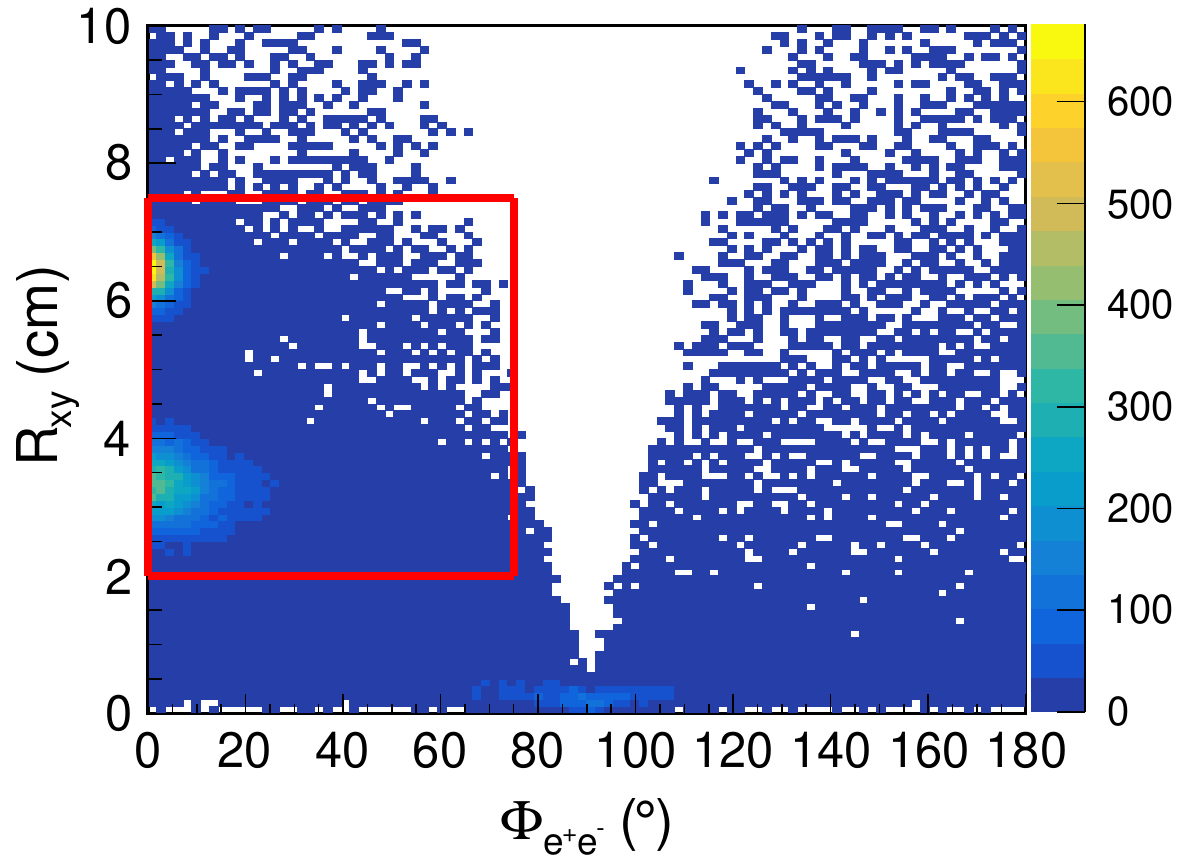}
		\put(47,60){$\bf (b)$}
	\end{overpic}
	\caption{Photon conversion veto criterion on the (a) $R_{xy}$ vs. $M^{\rm{BP}}(e^+e^-)$ and (b) $R_{xy}$ vs. $\Phi_{ee}$ plots from data. All events above or inside the red solid polygons are rejected.}	
	\label{scap_gamma_conversion}
\end{figure}

In order to reject photon conversion events and improve the signal-to-background ratio, two additional variables are introduced.
The first one is the invariant mass of the $e^+e^-$ pair at the beam pipe, $M^{\rm{BP}}(e^+e^-)$, which is determined by changing the reference point of the helices of all $e^{\pm}$ tracks to their respective points of intersection with the beam pipe, and subsequently recalculating the momentum vectors. 
This changes the direction of the vectors, but not their magnitudes.
The momenta of the $e^+e^-$ pair that is created at the beam pipe will instead be approximately parallel and the invariant mass close to the mass of two electrons.
The second variable is the opening angle of the $e^+e^-$ pair in the $x-y$ plane, $\Phi_{ee}$~\cite{PhysRevC.81.034911}.
For the $e^+e^-$ pairs that originate from the IP, such as $\eta'\rightarrow\pi^{+}\pi^{-}e^{+}e^{-}$ decays, the opening angles determined with the calculated momentum vectors are increased, and as a consequence, their invariant masses become larger than the true value. 
For conversion events, $\Phi_{ee}$ is expected to be close to zero.
In the two-dimensional distributions of $R_{xy}$ vs. $M^{\rm{BP}}(e^+e^-)$, and $R_{xy}$ vs. $\Phi_{ee}$, shown in Fig.~\ref{scap_gamma_conversion}, signal events and photon conversion events are well separated.
By selecting appropriate regions of these distributions, the photon conversion events are vetoed. 
First, we select all events to the high mass side of a curve defined by straight line segments between the points (0.004 GeV/$c^2$, 0 cm), (0.004 GeV/$c^2$, 2 cm), (0.03 GeV/$c^2$, 3 cm), and (0.07 GeV/$c^2$,10 cm) in the $R_{xy}$ vs. $M^{\rm{BP}}(e^+e^-)$ distribution. 
Then, we reject the events from $\Phi_{ee}<75^{\circ}$ when $2$ cm $<R_{xy}<7.5$ cm.
The effects of the photon conversion veto, have been estimated in MC simulations, and it removes 98.6\% of the $\eta'\rightarrow\pi^+\pi^-\gamma$ conversion background, while retaining $\sim$ 86\% of the $\eta'\rightarrow\pi^+\pi^-e^+e^-$ signal.

\begin{figure}[htbp]
	\centering
	\begin{overpic}[width=0.49\textwidth]{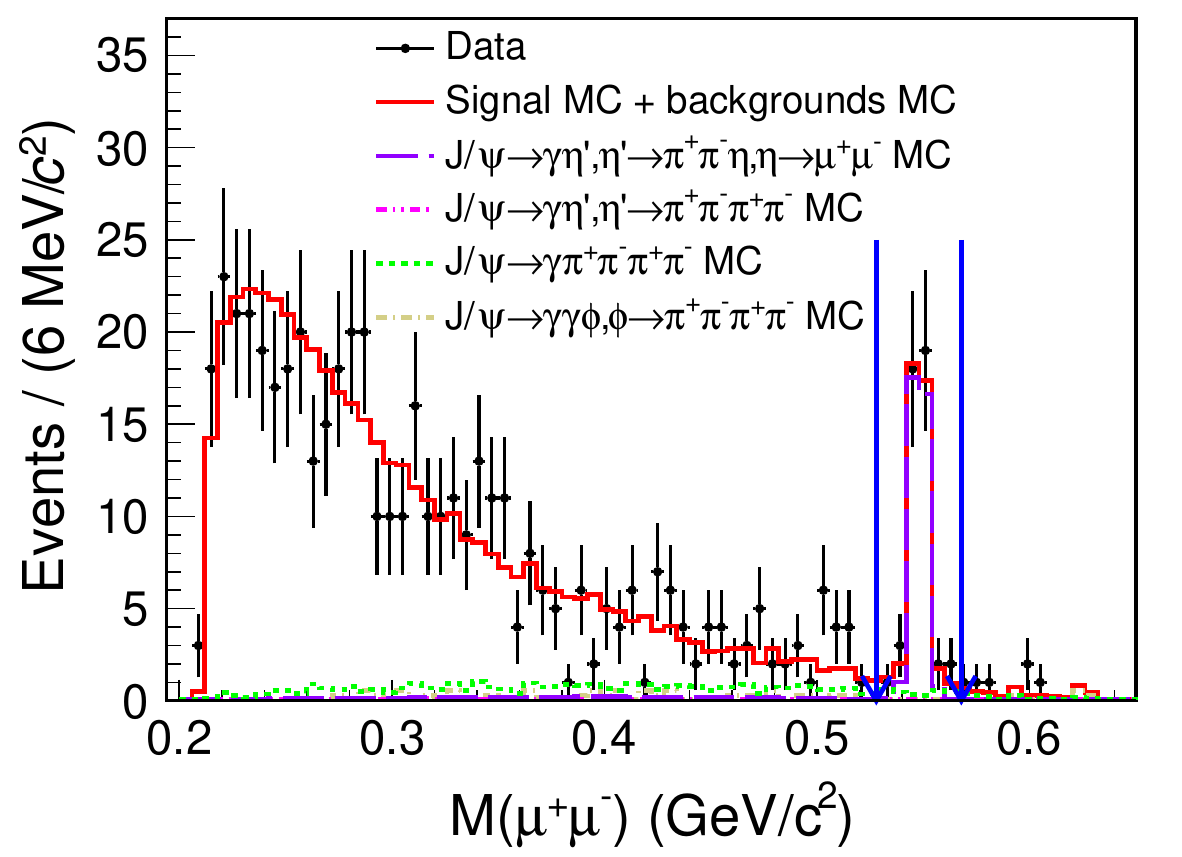}
	\end{overpic}
	\caption{The $\mu^{+}\mu^{-}$ invariant mass distribution for events satisfying $0.945\ \mathrm{GeV}/c^2<M(\pi^+\pi^-\mu^+\mu^-)<0.975\ \mathrm{GeV}/c^2$.
    The dots with error bars represent the data, the purple dashed histogram is the $J/\psi\rightarrow\gamma\eta', \eta'\rightarrow\pi^+\pi^-\eta, \eta\rightarrow\mu^+\mu^-$ MC shape, the pink dotted histogram is the $J/\psi\rightarrow\gamma\eta', \eta'\rightarrow\pi^+\pi^-\pi^+\pi^-$ MC shape, the green dashed histogram is the $J/\psi\rightarrow\gamma\pi^+\pi^-\pi^+\pi^-$ MC shape, the golden dotted histogram is the $J/\psi\rightarrow\gamma\gamma\phi, \phi\rightarrow\pi^+\pi^-\pi^+\pi^-$ MC shape, and the red solid histogram is the sum of the MC signals and all MC backgrounds.}	
	\label{Mpipi_mumu}
\end{figure}

For the $\eta'\rightarrow\pi^{+}\pi^{-}\mu^{+}\mu^{-}$ decay, $\chi^2_{\rm{4C}}(\gamma\pi^+\pi^-\mu^+\mu^-) < 25$ is required, and $\chi^2_{\rm{4C+ PID}}(\gamma\pi^+\pi^-\mu^+\mu^-)$ is required to be less than $\chi^2_{\rm{4C+ PID}}(\gamma\pi^+\pi^-\pi^+\pi^-)$ to suppress background from $J/\psi\rightarrow\gamma\pi^+\pi^-\pi^+\pi^-$.
The comparison of the $\mu^+\mu^-$ mass spectrum between data and MC simulation in Fig.~\ref{Mpipi_mumu}, where $0.945\ \mathrm{GeV}/c^2<M(\pi^+\pi^-\mu^+\mu^-)<0.975\ \mathrm{GeV}/c^2$ has been required, shows good agreement. To suppress the background from $\eta'\rightarrow\pi^{+}\pi^{-}\eta, \eta\rightarrow\mu^+\mu^-$, we require $|M(\mu^+\mu^-) - 0.548| > 0.02\ \mathrm{GeV}/c^2$.

Possible background events are analyzed with the same procedure using the inclusive MC sample, and they mainly originate from the background processes listed in Table~\ref{backgrounds}. 
For the dominant background channels, dedicated exclusive MC samples are generated to estimate their contributions to the $\pi^+\pi^-l^+l^-$ invariant mass spectrum in Fig.~\ref{fit_br}.
For the $\mathcal{B}$ determination, we require the fit range of $0.91\ \mathrm{GeV}/c^2<M(\pi^+\pi^-e^+e^-)<1.0\ \mathrm{GeV}/c^2$ and $0.9\ \mathrm{GeV}/c^2<M(\pi^+\pi^-\mu^+\mu^-)<1.0\ \mathrm{GeV}/c^2$ for the $\eta^\prime\rightarrow\pi^+\pi^-e^+e^-$ and $\eta^\prime\rightarrow\pi^+\pi^-\mu^+\mu^-$ decays, respectively.
In the measurements of the TFF, $\mathcal{A}_{CP}$ and search for ALPs, we further require $0.945\ \mathrm{GeV}/c^2<M(\pi^+\pi^-e^+e^-)<0.97\ \mathrm{GeV}/c^2$ and $0.945\ \mathrm{GeV}/c^2<M(\pi^+\pi^-\mu^+\mu^-)<0.975\ \mathrm{GeV}/c^2$ for the $\eta^\prime\rightarrow\pi^+\pi^-e^+e^-$ and $\eta^\prime\rightarrow\pi^+\pi^-\mu^+\mu^-$ decays, respectively. 
The aim is to minimize background interference and preserve the signals as much as possible.

\begin{table}[htbp]
	\small
	\renewcommand\arraystretch{1.1}
	\begin{center}
		\begin{tabular}{c|c|c|c}
        \hline
        \hline
        \multirow{2}*{Signal channel} & \multirow{2}*{Main background channels} & \multicolumn{2}{c}{Normalized events} \\
         \cline{3-4} ~ & & $\mathcal{B}$ & TFF/$\mathcal{A}_{CP}$/ALPs  \\ 
        \hline
         $\eta'\rightarrow\pi^+\pi^-e^+e^-$ & $J/\psi\rightarrow\gamma\eta', \eta'\rightarrow\gamma\pi^+\pi^-$                                 &$672\pm22$  &$598\pm20$\\
         \hline
         \multirow{5}*{$\eta'\rightarrow\pi^+\pi^-\mu^+\mu^-$} & $J/\psi\rightarrow\gamma\eta', \eta'\rightarrow\pi^+\pi^-\eta, \eta\rightarrow\mu^+\mu^-$        &$11\pm1$   &$9\pm1$\\
        & $J/\psi\rightarrow\gamma\eta', \eta'\rightarrow\pi^+\pi^-\pi^+\pi^-$                             &$206\pm2$  &$3\pm1$\\
        & $J/\psi\rightarrow\gamma\eta', \eta'\rightarrow\pi^+\pi^-\eta, \eta\rightarrow\gamma\pi^+\pi^-$  &$24\pm2$   &$-$\\
        & $ J/\psi\rightarrow\gamma\pi^+\pi^-\pi^+\pi^-$                                                   &$-$   &$36\pm11$\\
        & $J/\psi\rightarrow\gamma\gamma\phi, \phi\rightarrow\pi^+\pi^-\pi^+\pi^-$                                                                                 &$-$   &$10\pm2$\\
        \hline
        \hline
		\end{tabular}
	\end{center}
	\caption{Main background processes and their corresponding normalized events of the $\eta'\rightarrow\pi^{+}\pi^{-}l^{+}l^{-}$ decays. For the $\mathcal{B}$ determination, "$-$" indicates that the number of background events are left floating in the fit. 
	In the measurements of the TFF, $\mathcal{A}_{CP}$ and search for ALPs, "$-$" indicates that the background channel is disregarded due to the extremely low number of background events in the signal mass range.}
	\label{backgrounds}
\end{table}	

\section{BRANCHING FRACTION MEASUREMENTS}
To determine the numbers of the $\eta^\prime\rightarrow\pi^+\pi^-l^+l^-$ events, an unbinned maximum likelihood fits are performed to the $\pi^+\pi^-l^+l^-$ invariant mass spectra. 
The branching fractions of $\eta'\rightarrow\pi^+\pi^-l^+l^-$ are determined by
\begin{equation}\label{br}
\mathcal{B}(\eta'\rightarrow\pi^+\pi^-l^+l^-)=\frac{N_{\rm{sig}}}{N_{J/\psi}\times \mathcal{B}(J/\psi\rightarrow\gamma\eta')\times \varepsilon},
\end{equation}
where $N_{\rm{sig}}$ are the signal yields determined from data, $\varepsilon$ are the detection efficiencies for the decays $\eta^\prime\rightarrow\pi^+\pi^-l^+l^-$, which are determined from the simulated samples, $\mathcal{B}(J/\psi\rightarrow\gamma\eta')$ is the branching fraction of $J/\psi\rightarrow\gamma\eta'$, $(5.21\pm0.17)\times10^{-3}$\cite{PDG_2022} and $N_{J/\psi}$ is the number of $J/\psi$ events, $(10087\pm44)\times10^{6}$~\cite{BESIII_2021_njpsi}. 

\begin{figure}[htbp]
	\centering
	\begin{overpic}[width=0.49\textwidth]{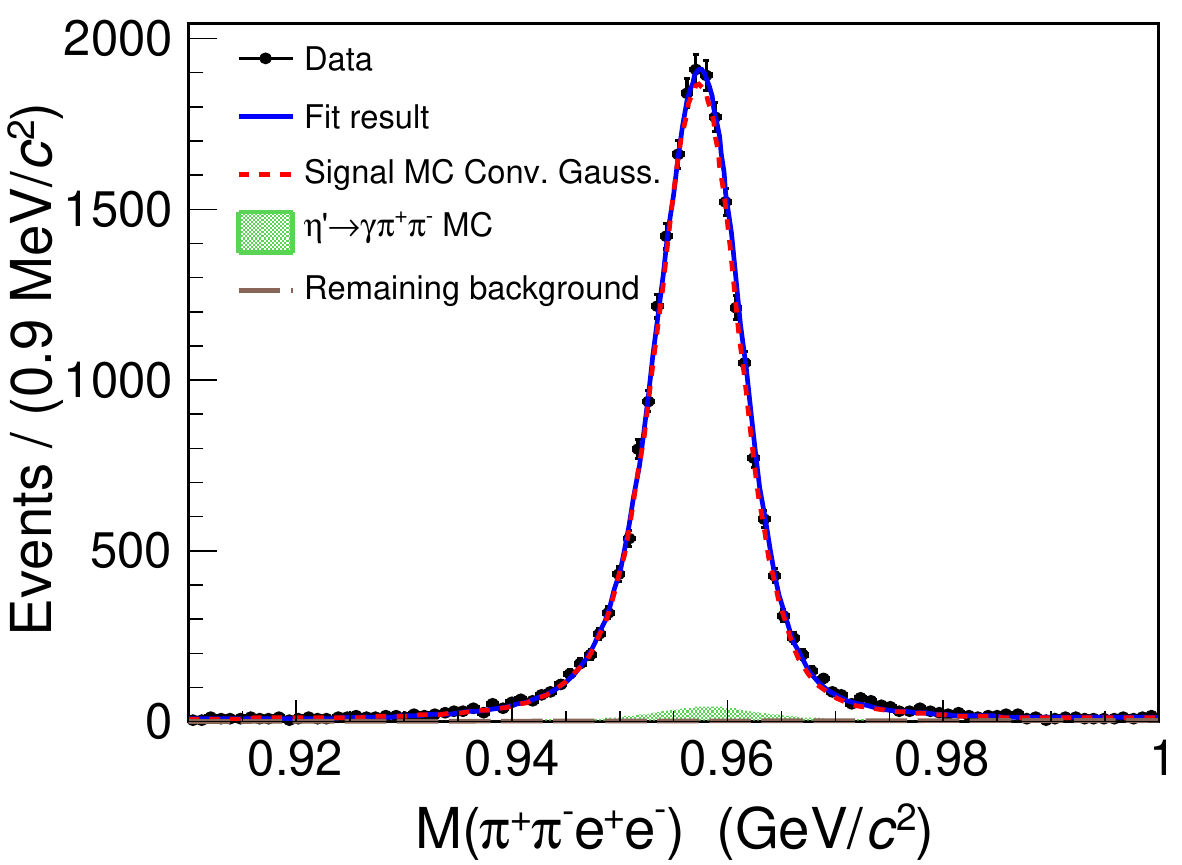}
		\put(80,60){$\bf (a)$}
	\end{overpic}
	\begin{overpic}[width=0.49\textwidth]{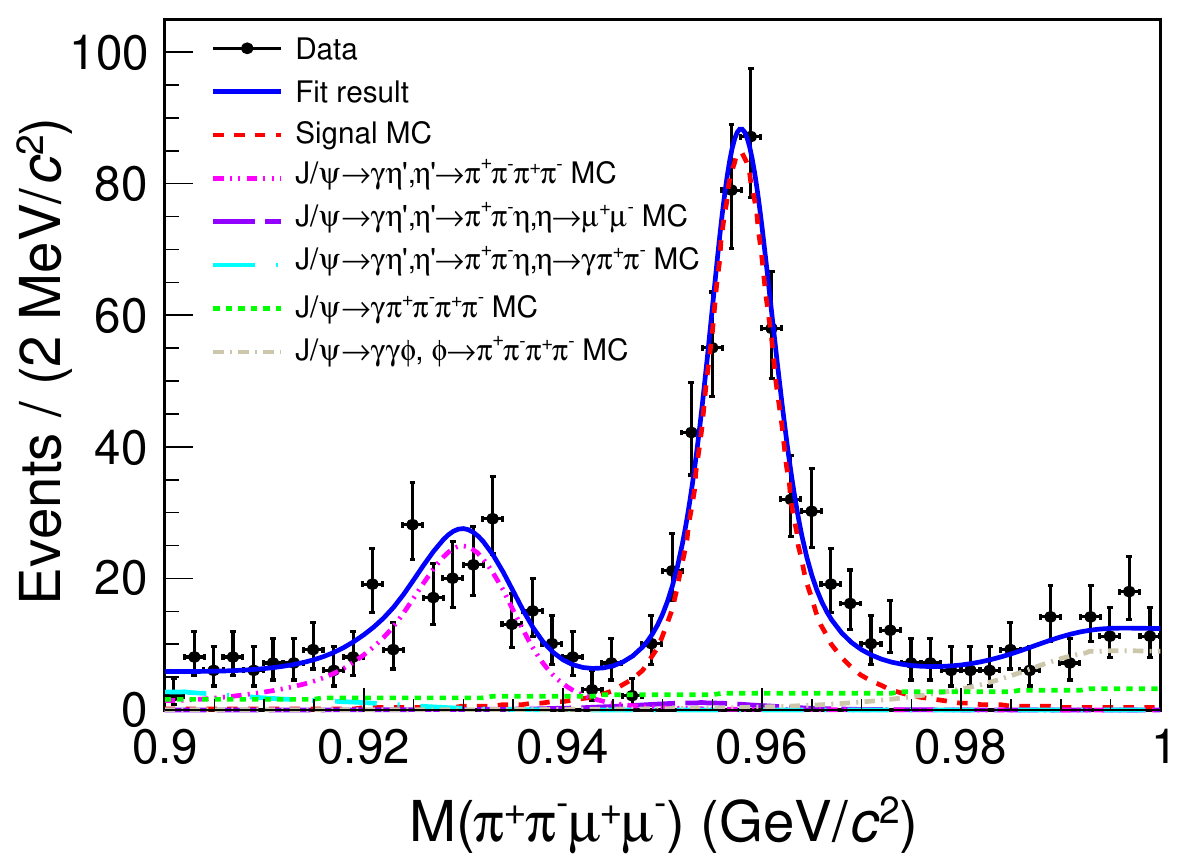}
		\put(80,60){$\bf (b)$}
	\end{overpic}
	\caption{Fits to the invariant mass distributions of (a) $\pi^+\pi^-e^+e^-$ and (b) $\pi^+\pi^-\mu^+\mu^-$.
    The dots with error bars represent the data, and the blue solid lines are the total fit result.
    For the $\eta'\rightarrow\pi^+\pi^-e^+e^-$ decay, the red dashed histogram represents the MC signal shape convolved with a Gaussian function, the green area is the $J/\psi\rightarrow\gamma\eta', \eta'\rightarrow\gamma\pi^+\pi^-$ MC shape, and the brown dashed histogram represents the remaining background described by the first-order Chebychev polynomial.
    For the $\eta'\rightarrow\pi^+\pi^-\mu^+\mu^-$ decay, the red dashed histogram represents the MC signal shape, the pink dotted histogram is the $J/\psi\rightarrow\gamma\eta', \eta'\rightarrow\pi^+\pi^-\pi^+\pi^-$ MC shape, the purple dashed histogram is the $J/\psi\rightarrow\gamma\eta', \eta'\rightarrow\pi^+\pi^-\eta, \eta\rightarrow\mu^+\mu^-$ MC shape. The light blue dotted histogram is the $J/\psi\rightarrow\gamma\eta', \eta'\rightarrow\pi^+\pi^-\eta, \eta\rightarrow\gamma\pi^+\pi^-$ MC shape, the green dashed histogram is the $J/\psi\rightarrow\gamma\pi^+\pi^-\pi^+\pi^-$ MC shape, and the golden dotted histogram is the $J/\psi\rightarrow\gamma\gamma\phi,\phi\rightarrow\pi^{+}\pi^{-}\pi^{+}\pi^{-}$ MC shape.}	
	\label{fit_br}
\end{figure}

For the $\eta'\rightarrow\pi^+\pi^-e^+e^-$ decay fit, the signal is represented by the MC shape convolved with a Gaussian function to account for the difference in resolution between data and MC simulation.
The peaking background arising from the photon conversion process $\eta'\rightarrow\gamma\pi^+\pi^-$ is described by the MC simulation shape,
and the number of background events, which are listed in Table~\ref{backgrounds}, are fixed according to the branching fractions from the PDG~\cite{PDG_2022}.
Combinatorial background is represented by a first-order Chebychev polynomial.
The fit result shown in Fig.~\ref{fit_br} (a) yields $22725\pm155$ signal events.
The goodness of fit is studied using a $\chi^{2}$ test and the $\chi^{2}$ value per number of degrees of freedom ($ndf$) is found to be $\chi^{2}/ndf=95.4/95.0$.
With a detection efficiency of $(17.49\pm0.04)\%$, the branching fraction of $\eta'\rightarrow\pi^+\pi^-e^+e^-$ is calculated to be $(2.45\pm0.02) \times 10^{-3}$, where the uncertainties are statistical only.

For the $\eta'\rightarrow\pi^{+}\pi^{-}\mu^{+}\mu^{-}$ decay fit, the shapes of signal and $J/\psi\rightarrow\gamma\pi^+\pi^-\pi^+\pi^-$ are taken from the MC simulation.
From a topological analysis, possible backgrounds arise from the $J/\psi\rightarrow\gamma\gamma\phi,\phi\rightarrow\pi^{+}\pi^{-}\pi^{+}\pi^{-}$ decay in the $\eta'$ mass distribution around 1 GeV/$c^{2}$. 
A MC sample of $J/\psi\rightarrow\gamma\gamma\phi,\phi\rightarrow\pi^{+}\pi^{-}\pi^{+}\pi^{-}$ is generated to describe these backgrounds.
The magnitude of the $J/\psi\rightarrow\gamma\gamma\phi,\phi\rightarrow\pi^{+}\pi^{-}\pi^{+}\pi^{-}$ and $J/\psi\rightarrow\gamma\pi^+\pi^-\pi^+\pi^-$ backgrounds are determined from the fit. 
All other background events, which are listed in Table~\ref{backgrounds}, are fixed according to the branching fractions from the PDG~\cite{PDG_2022}.
The fit result shown in Fig.~\ref{fit_br} (b) yields $434\pm25$ signal events, and the goodness of fit is found to be $\chi^{2}/ndf=88.9/97.0$.
With a detection efficiency of $(37.95\pm0.07)\%$, the branching fraction of $\eta'\rightarrow\pi^+\pi^-\mu^+\mu^-$ is calculated to be $(2.16\pm0.12) \times 10^{-5}$, where the uncertainties are statistical only.

\section{TRANSITION FORM FACTOR MEASUREMENTS}
\subsection{VMD Model}
\begin{figure}[htbp]
	\centering
	\begin{overpic}[width=0.70\textwidth]{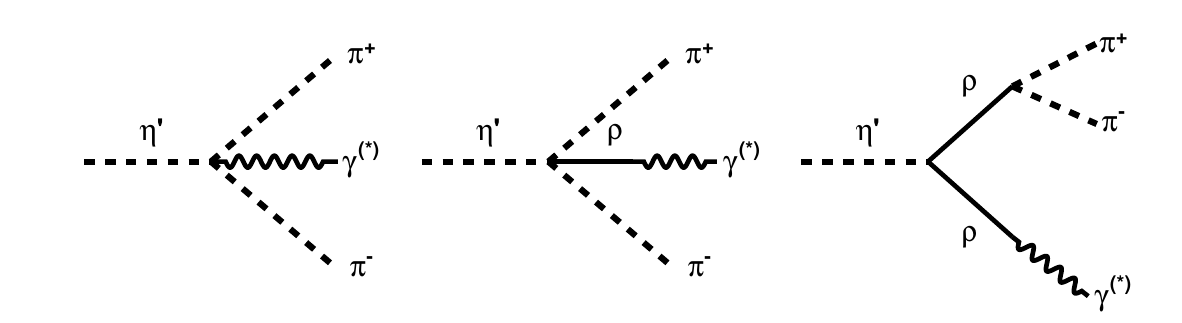}
	\end{overpic}
	\caption{Contact term corresponding to the axial anomaly (left) and other VMD contributions (middle and right) to the decay $\eta'\rightarrow\pi^+\pi^-\gamma^{\ast}$~\cite{1010_2378}.}	
	\label{box}
\end{figure}

The $\eta'\rightarrow\pi^{+}\pi^{-}l^{+}l^{-}$ decays are similar to $\eta'\rightarrow\gamma\pi^{+}\pi^{-}$, in which the photon is replaced by an off-shell one that decays into a $l^{+}l^{-}$ pair. 
Within the VMD model, it is factorized into three separate parts~\cite{1010_2378}, as illustrated in Fig.~\ref{box}. The contact term represents the contribution from the axial anomaly, and the other two diagrams give the VMD interactions where the pseudoscalar does not interact directly with the photon.

The four-momenta for the decays
$\eta'(P)\rightarrow\pi^{+}(p_+)\pi^{-}(p_-)l^{+}(k_+)l^{-}(k_-)$ are defined as $P=p_++p_-+k_++k_-$~\cite{KL_pipiee_2}. The relevant variables are written as
\begin{equation}\label{VMDF1}
s_{\pi\pi}=(p_++p_-)^2,
s_{ll}=(k_++k_-)^2.
\end{equation}
\begin{equation}\label{VMDF2}
\beta_{\pi}=\sqrt{1-\frac{4M^2(\pi^+\pi^-)}{s_{\pi\pi}}},
\beta_{l}=\sqrt{1-\frac{4M^2(l^+l^-)}{s_{ll}}}.
\end{equation}
The squared decay amplitude is given as~\cite{Event_generators_2012_etap_pipill}
\begin{equation}
\begin{aligned}
\overline{\vert \mathcal{A}_{\eta'\rightarrow\pi^+\pi^-l^+l^-} \vert}^2(s_{\pi\pi}, s_{ll}, \theta_{\pi}, \theta_{l}, \varphi)
&=\frac{e^2}{8k^2}\vert \mathcal{M}(s_{\pi\pi}, s_{ll}) \vert^2 \times \lambda(M^2(\eta'), s_{\pi\pi}, s_{ll}) \\
&\times [1-\beta^2_l\sin^2{\theta_l}\sin^2{\varphi}]s_{\pi\pi}\beta^2_\pi\sin^2{\theta_\pi},
\label{amp}
\end{aligned}
\end{equation}
to the leading order with only the magnetic term. The high order corrections with the electric term and the mixing term are visibly smaller than the leading one~\cite{1010_2378}.
Here $\theta_\pi$ is the polar angle of the $p_+$ in the $p_+p_-$ rest frame with respect to the flight direction of the $p_+p_-$ in the $P$ rest frame, $\theta_l$ is the polar angle of the $k_-$ in the $k_+k_-$ rest frame with respect to the flight direction of the $k_+k_-$ in the $P$ rest frame, and $\varphi$ is the azimuthal angle between the plane formed by $p_+p_-$ in the $P$ rest frame and the corresponding plane formed by the $k_+k_-$~\cite{amp_k1k2_1965_1, amp_k1k2_1860_2}. The $\mathrm{K\Ddot{a}ll\acute{e}n}$ function $\lambda$ is defined as $\lambda(a,b,c) = a^2 + b^2 + c^2-2(ab + bc + ca)$.
The magnetic form factor $\mathcal{M}(s_{\pi\pi}, s_{ll})$ is given by
\begin{equation}
\mathcal{M}(s_{\pi\pi}, s_{ll})=\mathcal{M}_{\mathrm{mix}} \times \mathrm{VMD}(s_{\pi\pi}, s_{ll}),
\end{equation}
where $\mathcal{M}_{\mathrm{mix}}$ is the pseudoscalar mesons mixing parameter~\cite{1010_2378}, VMD$(s_{\pi\pi}, s_{ll})$ is the VMD factor~\cite{1010_2378, vmd_factor_1993_1} and derived from the VMD model with finite-width corrections as~\cite{vmd_factor_1968_2}
\begin{equation}
\begin{aligned}
\mathrm{VMD}(s_{\pi\pi}, s_{ll}) = 
&1 - \frac{3}{4}(c_1-c_2+c_3) + \frac{3}{4}(c_1-c_2-c_3) \times \frac{m^2_V}{m^2_V-s_{ll}-im_V\Gamma(s_{ll})} \\
&+ \frac{3}{2}c_3\frac{m^2_{V,\pi}}{m^2_{V,\pi}-s_{\pi\pi}-im^2_{V,\pi}\Gamma(s_{\pi\pi})}\times \frac{m^2_V}{m^2_V-s_{ll}-im_V\Gamma(s_{ll})}.
\label{amp_vmd}
\end{aligned}
\end{equation}
Here $m_V$ and $m_{V,\pi}$ are the masses of the vector meson $V(V\rightarrow l^{+}l^{-})$ and $V_{\pi}(V_{\pi}\rightarrow\pi^{+}\pi^{-})$, respectively.
The parameters $c_{1-3}$ are model dependent and determine the contributions of the different interaction terms illustrated in Fig.~\ref{box}.
$\Gamma$ is its total width~\cite{Event_generators_2012_etap_pipill} as
\begin{align}
    \Gamma(s)=g\left(\frac{\sqrt{s}}{m_{V,i}}\right)\left(\frac{1-\frac{4m^2_{i}}{s}}{1-\frac{4m^2_{i}}{m^2_{V,i}}}\right)^{\frac{3}{2}} \Theta(s-4m^2_{i})\ (i=l\ \mathrm{or}\ \pi;\ V,i=V\ \mathrm{or}\ V,\pi),
\end{align}
where $g = 149.1$ MeV, and $\Theta$ is the Heaviside step function. 

\subsection{Analysis Method and Background Treatment}
\label{sec:TFF mathod}
Unbinned maximum likelihood fits to the $e^{+}e^{-}$ and $\pi^{+}\pi^{-}$ mass spectra are performed to determine the fit parameters in the VMD factor given in Eq.~\ref{amp_vmd}.
The probability-density function (PDF) to observe the $i$-th event characterized by the measured four-momenta $\xi_i$ of the particles in the final state is
\begin{equation}
\mathcal{P}(\xi_i)=\frac{\vert\mathcal{A}(\xi_i)\vert^2\varepsilon(\xi_i)}{\int\vert\mathcal{A}(\xi)\vert^2\varepsilon(\xi)\mathrm{d}\xi},
\end{equation}
where $\mathcal{A}$ is the amplitude as shown in Eq. \ref{amp}, and $\varepsilon(\xi_i)$ is the detection efficiency.
The fit minimizes the  negative log-likelihood value
\begin{equation}
-\ln{\mathcal{L}} = -\omega'[\sum^{N_{\mathrm{data}}}_{i=1}\ln{\mathcal{P}(\xi_i)}-\omega_{\mathrm{bkg1}}\sum^{N_{\mathrm{bkg1}}}_{j=1}\ln{\mathcal{P}(\xi_j)}-\omega_{\mathrm{bkg2}}\sum^{N_{\mathrm{bkg2}}}_{k=1}\ln{\mathcal{P}(\xi_k)}-...],
\end{equation}
where $i$ run over all accepted data, and $j$, $k$, ... run over the other considered background events.
There corresponding number of events are denoted by $N_{\mathrm{data}}$, $N_{\mathrm{bkg1}}$ and $N_{\mathrm{bkg2}}$.
$\omega_{\mathrm{bkg1}}=\frac{N'_{\mathrm{bkg1}}}{N_{\mathrm{bkg1}}}$ and $\omega_{\mathrm{bkg2}}=\frac{N'_{\mathrm{bkg2}}}{N_{\mathrm{bkg2}}}$ are the weights of the backgrounds, where $N'_{\mathrm{bkg1}}$ and $N'_{\mathrm{bkg2}}$ are their contributions according to branching fractions taken from PDG.
To obtain an unbiased uncertainty estimation, the normalization factor $\omega'$~\cite{Langenbruch:2019nwe} is considered, described as
\begin{equation}
\omega'=\frac{N_{\mathrm{data}}-N_{\mathrm{bkg1}}\omega_{\mathrm{bkg1}}-N_{\mathrm{bkg2}}\omega_{\mathrm{bkg2}}}{N_{\mathrm{data}}+N_{\mathrm{bkg1}}\omega^{2}_{\mathrm{bkg1}}+N_{\mathrm{bkg2}}\omega^{2}_{\mathrm{bkg2}}}.
\end{equation}

For the $\eta'\rightarrow\pi^+\pi^-e^+e^-$ decay, the number of the background events of $J/\psi\rightarrow\gamma\eta', \eta'\rightarrow\gamma\pi^+\pi^-$ is fixed according to the branching fractions from the PDG~\cite{PDG_2022}, and the non-peaking background distribution is estimated by using the $\eta'$ one-dimensional sideband events.
The sideband is defined as $M(\pi^+\pi^-e^+e^-)\in[0.915, 0.9275]\bigcup[0.985, 0.9975]\ \mathrm{GeV}/c^{2}$.
For the $\eta'\rightarrow\pi^+\pi^-\mu^+\mu^-$ decay, the number of the background events of $J/\psi\rightarrow\gamma\pi^+\pi^-\pi^+\pi^-$ is obtained by fitting the mass spectrum of $\pi^+\pi^-\mu^+\mu^-$, and other backgrounds yields are fixed according to the branching fractions from the PDG~\cite{PDG_2022}.
All the numbers of the background events, which are listed in Table~\ref{backgrounds}, are obtained in the $\eta'$ mass regions, $0.945\ \mathrm{GeV}/c^{2} < M(\pi^{+}\pi^{-}e^{+}e^{-}) < 0.97\ \mathrm{GeV}/c^{2}$ and $0.945\ \mathrm{GeV}/c^{2} < M(\pi^{+}\pi^{-}\mu^{+}\mu^{-}) < 0.975\ \mathrm{GeV}/c^{2}$.

\subsection{Fit Results}
\label{sec:TFF results}
By adjusting the values of the $c_i$-parameters~\cite{ci_2010}, we can switch between various VMD models:
\begin{itemize}
    \item Hidden gauge model (Model I): $c_1 - c_2 = c_3 = 1$;
    \item Full VMD model (Model II): $c_1 - c_2 = 1/3, c_3 = 1$;
    \item Modified VMD model (Model III): $c_1 - c_2 \neq c_3$.
\end{itemize}
Since the dominant contribution is from $\rho^0$ and its width is large, the parameterization of its shape plays a vital role in describing the $e^+e^-$ and $\pi^+\pi^-$ mass spectra. 
For $\eta'\rightarrow\pi^+\pi^-e^+e^-$, the $\rho^0$ contribution is insufficient to describe the $\pi^+\pi^-$ mass spectra of data and an extra contributions from $\omega$ is necessary.
If the $\omega\rightarrow\pi^+\pi^-$ decay is taken into account, the propagator is written as
\begin{equation}
\frac{m^2_{V,\pi}}{m^2_{V,\pi}-s_{\pi\pi}-im^2_{V,\pi}\Gamma(s_{\pi\pi})}+\beta e^{i\theta}\frac{m^2_{\omega}}{m^2_{\omega}-s_{\pi\pi}-im^2_{\omega}\Gamma_{\omega}},
\end{equation}
where $\beta e^{i\theta}$ is the complex coupling parameter, and $\Gamma_{\omega} = 8.68\ \rm{MeV}$, which is taken from the PDG~\cite{PDG_2022}.

\begin{figure}[htbp]
	\centering
	\begin{overpic}[width=0.49\textwidth]{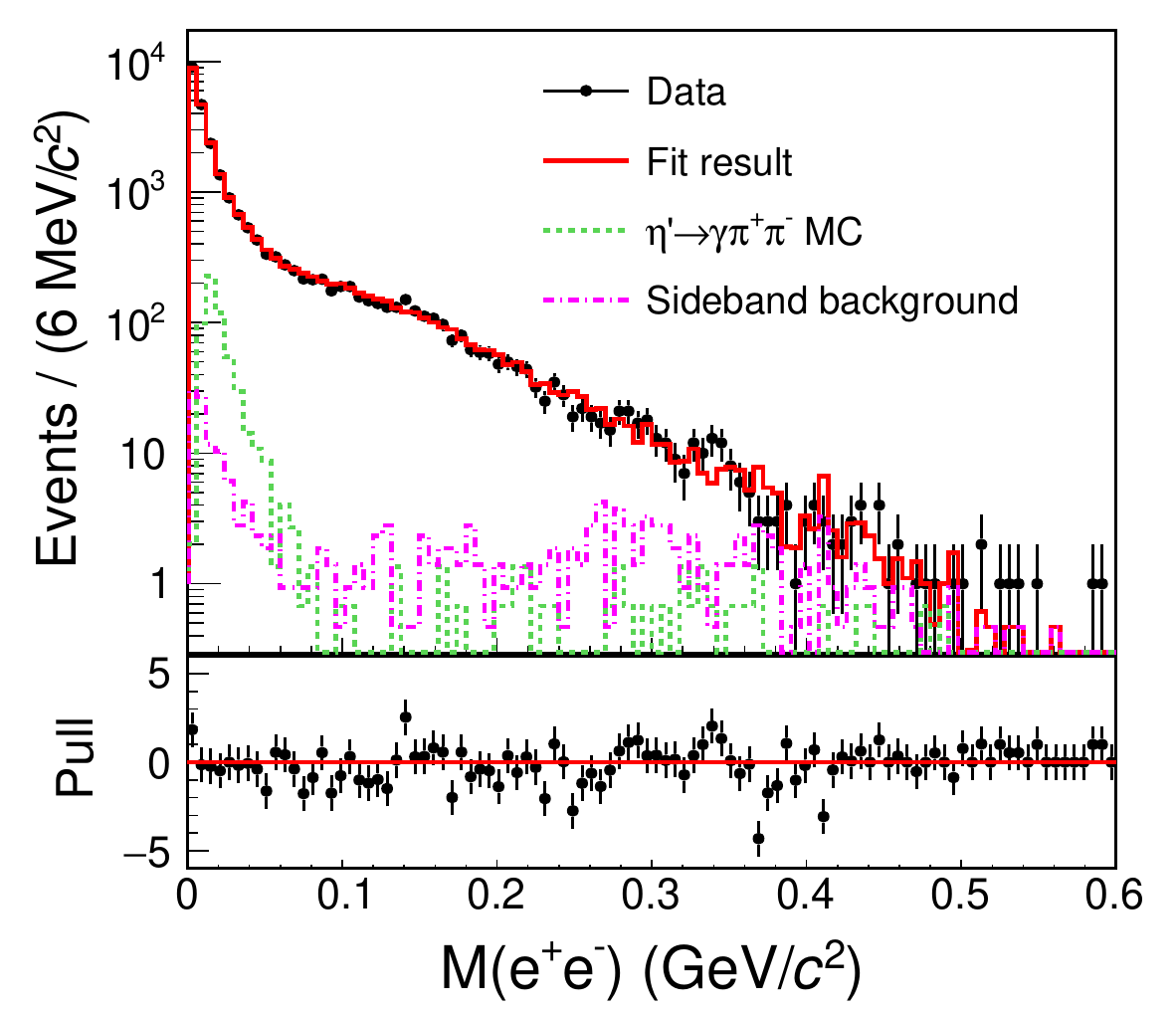}
	\put(85,76){{(a)}}
	\end{overpic}
	\begin{overpic}[width=0.49\textwidth]{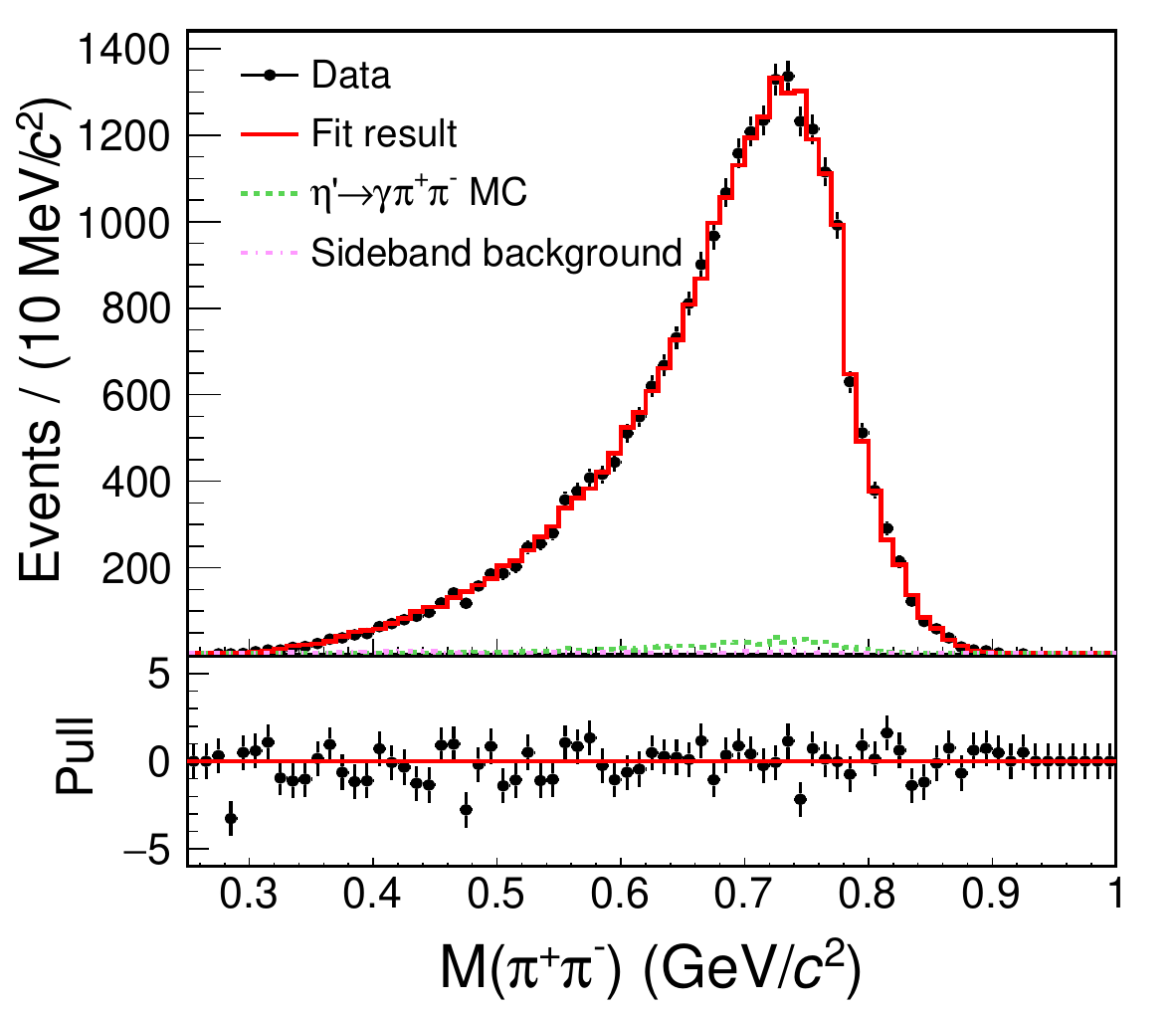}
	\put(85,76){{(b)}}
	\end{overpic}
	\caption{Fits to the invariant mass distributions of (a) $e^+e^-$ and (b) $\pi^+\pi^-$ for $\eta'\rightarrow\pi^+\pi^-e^+e^-$ with $c_1-c_2 = c_3 = 1$ (Model I). 
    The dots with error bars represent data, and the red solid histogram are the total fit results. The green dashed histogram are the $J/\psi\rightarrow\gamma\eta', \eta'\rightarrow\gamma\pi^{+}\pi^{-}$ MC shapes, and the pink dotted histogram are the backgrounds obtained from the $\eta'$ sideband.}	
	\label{TFF_fit_e}
\end{figure}

\begin{figure}[htbp]
	\centering
	\begin{overpic}[width=0.49\textwidth]{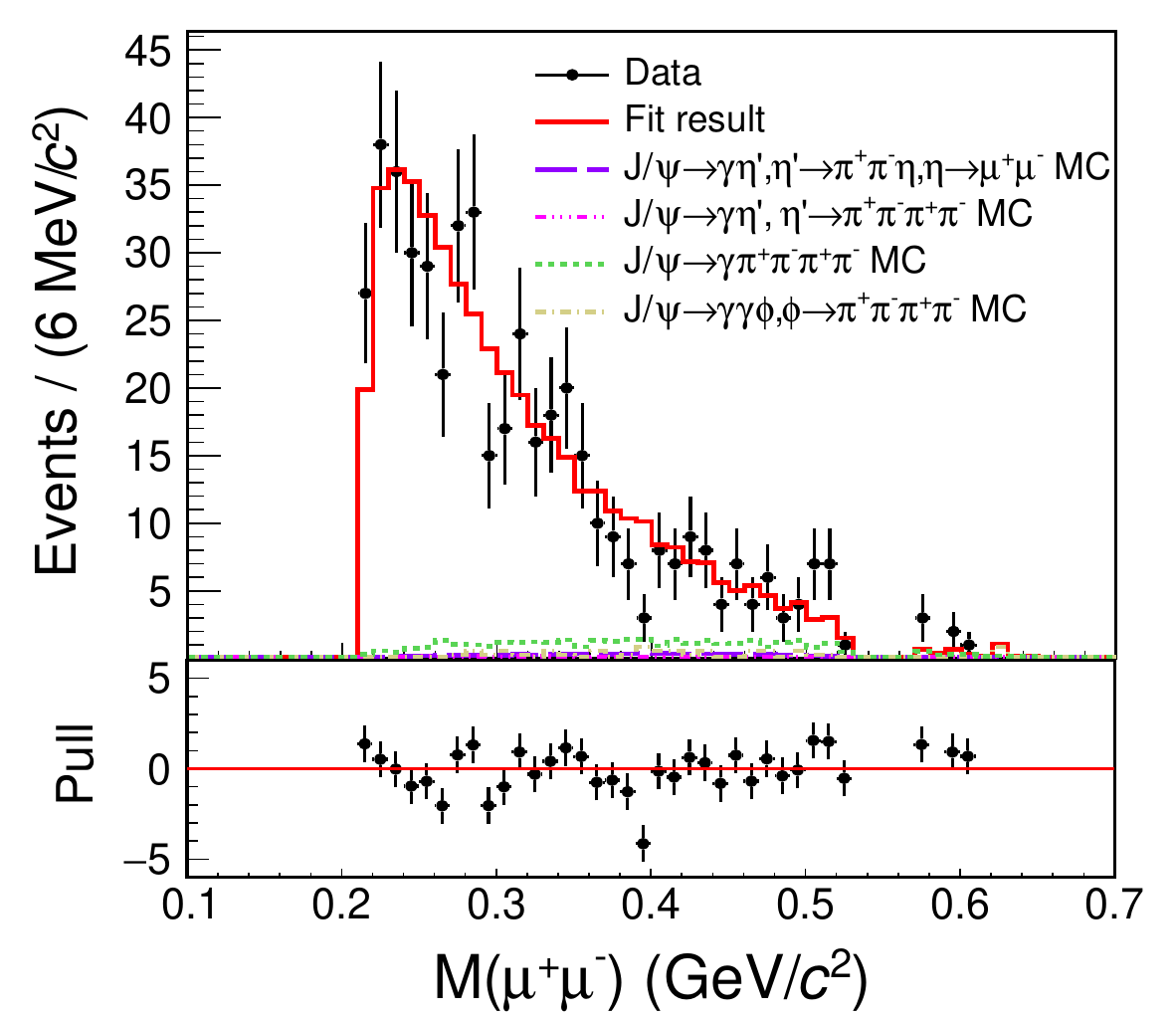}
	\put(85,76){{(a)}}
	\end{overpic}
	\begin{overpic}[width=0.49\textwidth]{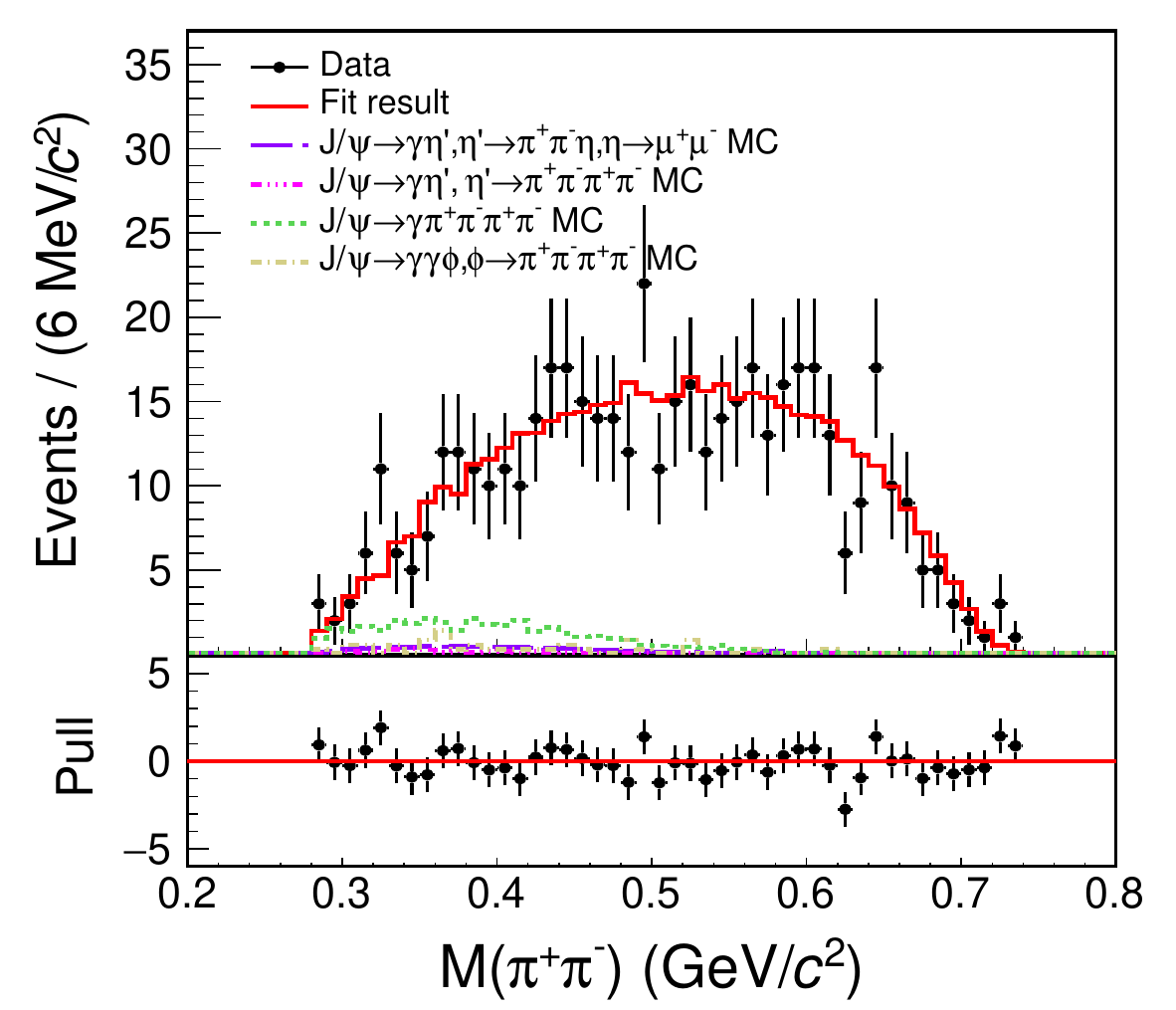}
	\put(85,76){{(b)}}
	\end{overpic}
	\caption{Fits to the invariant mass distributions of (a) $\mu^+\mu^-$ and (b) $\pi^+\pi^-$ for $\eta'\rightarrow\pi^+\pi^-\mu^+\mu^-$ with $c_1-c_2 = c_3 = 1$ (Model I). 
    The dots with error bars represent data, and the red solid histogram are the total fit result. The purple dashed histogram are the $J/\psi\rightarrow\gamma\eta', \eta'\rightarrow\pi^+\pi^-\eta, \eta\rightarrow\mu^+\mu^-$ MC shapes, the pink dotted histogram are the $J/\psi\rightarrow\gamma\eta', \eta'\rightarrow\pi^+\pi^-\pi^+\pi^-$ MC shapes, the green dashed histogram are the $J/\psi\rightarrow\gamma\pi^+\pi^-\pi^+\pi^-$ MC shapes, and the golden dotted histogram are the sums of the other background MC shapes.}	
	\label{TFF_fit_mu}
\end{figure}

The fit results of the three models mentioned above are summarized in Table~\ref{TFF_pipiee} and Table~\ref{TFF_pipimumu}, while those of Model I are shown in Fig.~\ref{TFF_fit_e} and Fig.~\ref{TFF_fit_mu}.
Based on Model I, we calculate a weighted average of $b_{\eta'}=1.30\pm0.19\ (\mathrm{GeV}/c^{2})^{-2}$ for $\eta'\rightarrow\pi^{+}\pi^{-}e^{+}e^{-}$ and $\eta'\rightarrow\pi^{+}\pi^{-}\mu^{+}\mu^{-}$ combined, where the uncertainty is obtained by combining statistical and systematic uncertainties. 
A test is also made with $c_1$, $c_2$, $c_3$ floating but with $c_1 - c_2 = c_3$~\cite{ci_2010}. 
The fit to our data gives $c_1 - c_2 = c_3 = 1.03 \pm 0.03$, which is consistent with the assumption of the hidden gauge model. 
Due to the limited statistics in the high $e^{+}e^{-}$ mass region, the statistical uncertainties of $m_{V}$ for the three models and $c_1 - c_2$ for Model III are relatively large.
At present, the fit results can not distinguish between the models.

\begin{table}[htbp]
	\small
	\renewcommand\arraystretch{1.1}
	\begin{center}
		\begin{tabular}{c|c|c|c}
			\hline
			\hline
        \multirow{2}*{$\eta'\rightarrow\pi^+\pi^-e^+e^-$} & Model I & Model II & Model III \\
         \cline{2-4} ~  & $c_1 - c_2 = c_3 = 1$ & $c_1 - c_2 = 1/3, c_3 = 1$ & $c_1 - c_2 \neq c_3$ \\
        \hline
        $m_V (\mathrm{MeV}/c^{2})$        & $954.3\pm87.8\pm36.4$ & $857.4\pm76.5$ & $787.5\pm173.9$  \\
        $m_{V,\pi} (\mathrm{MeV}/c^{2})$  & $765.3\pm1.2\pm20.2$  & $765.4\pm1.2$  & $764.8\pm1.3$  \\
        $m_{\omega} (\mathrm{MeV}/c^{2})$ & $778.7\pm1.3\pm17.3$  & $778.7\pm1.3$  & $778.7\pm1.4$  \\
        $\beta(10^{-3})$       & $8.5\pm1.4\pm0.7$     & $8.5\pm1.4$    & $8.1\pm1.5$  \\
        $\theta$               & $1.4\pm0.3\pm0.1$     & $1.4\pm0.3$    & $1.4\pm0.3$  \\
        $c_1-c_2$              & $1$ & $1/3$ & $-0.03\pm1.09$  \\
        $c_3$                  & $1$ & $1$ & $1.03\pm0.03$  \\
        $\chi^{2}/ndof (e^{+}e^{-}, \pi^{+}\pi^{-})$ &77.9/82.0, 47.8/65.0   &78.7/82.0, 47.6/65.0  &79.4/82.0, 45.1/65.0\\
        \hline
        $b_{\eta'} (\mathrm{GeV}/c^{2})^{-2}$ & $1.10\pm0.20\pm0.07$ & $1.36\pm0.24$ & $1.61\pm0.71$  \\
			\hline
			\hline
		\end{tabular}
	\end{center}
	\caption{The TFF fit results of the $\eta'\rightarrow\pi^+\pi^-e^+e^-$ decay.}
	\label{TFF_pipiee}
\end{table}	

\begin{table}[htbp]
	\small
	\renewcommand\arraystretch{1.1}
	\begin{center}
		\begin{tabular}{c|c|c|c}
			\hline
			\hline
        \multirow{2}*{$\eta'\rightarrow\pi^+\pi^-\mu^+\mu^-$} & Model I & Model II & Model III \\
        \cline{2-4} ~  & $c_1 - c_2 = c_3 = 1$ & $c_1 - c_2 = 1/3, c_3 = 1$ & $c_1 - c_2 \neq c_3$ \\
        \hline
        $m_V (\mathrm{MeV}/c^{2})$ & $649.4\pm55.9\pm35.6$ & $601.6\pm25.7$ & $589.6\pm25.9$\\
        $m_{V,\pi} (\mathrm{MeV}/c^{2})$ & $757.3\pm24.1\pm18.0$ & $765.4\pm18.8$ & $774.4\pm43.5$\\
        $c_1-c_2$ & $1$ & $1/3$ & $0.01\pm0.45$ \\
        $c_3$     & $1$ & $1$   & $0.98\pm0.40$ \\
        $\chi^{2}/ndof (\mu^{+}\mu^{-}, \pi^{+}\pi^{-})$ &48.1/34.0, 32.9/46.0   &48.3/34.0, 32.9/46.0  &49.7/35.0, 32.4/46.0\\
        \hline
        $b_{\eta'} (\mathrm{GeV}/c^{2})^{-2}$ & $2.37\pm0.41\pm0.27$ & $2.76\pm0.24$ & $2.88\pm0.25$\\
			\hline
			\hline
		\end{tabular}
	\end{center}
	\caption{The TFF fit results of the $\eta'\rightarrow\pi^+\pi^-\mu^+\mu^-$ decay.}
	\label{TFF_pipimumu}
\end{table}

\section{$CP$-VIOLATING ASYMMETRY}
\begin{figure}[htbp]
	\centering
	\begin{overpic}[width=0.49\textwidth]{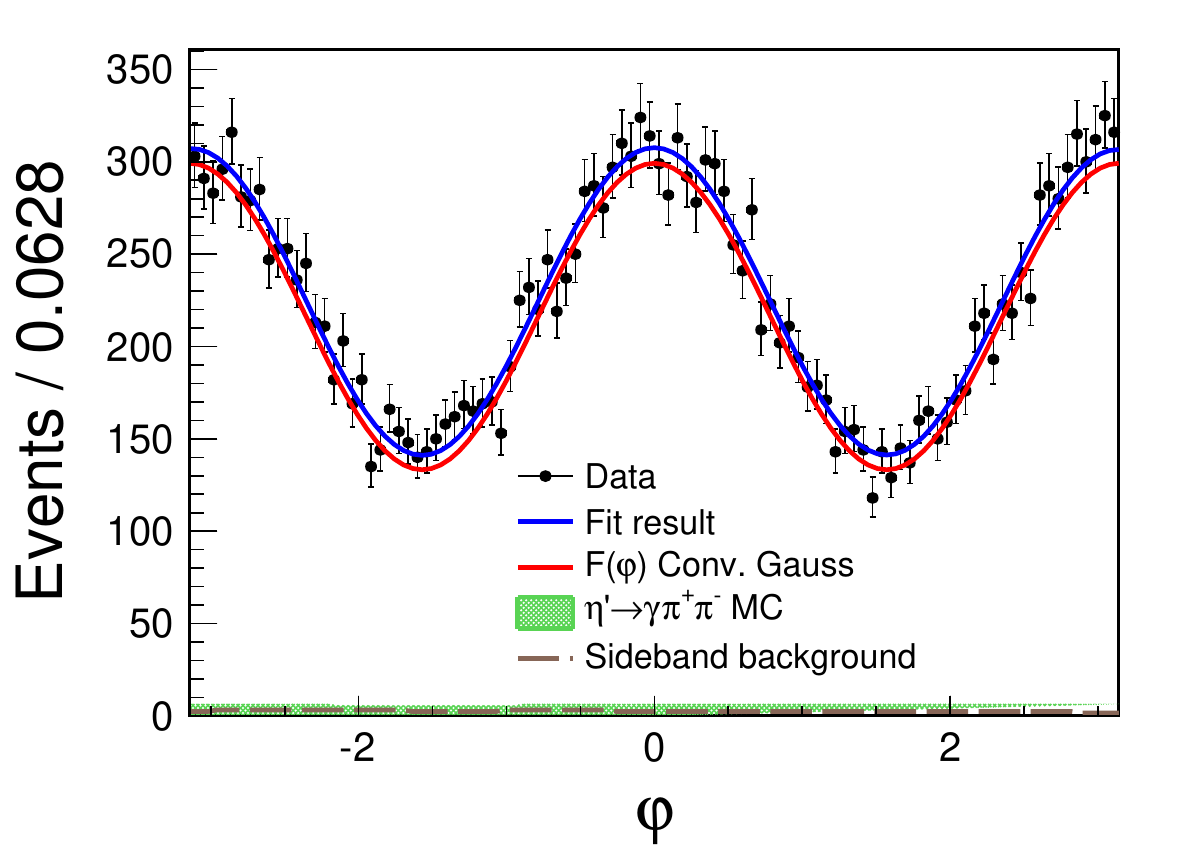}
	\put(75,60){{(a)}}
	\end{overpic}
	\begin{overpic}[width=0.49\textwidth]{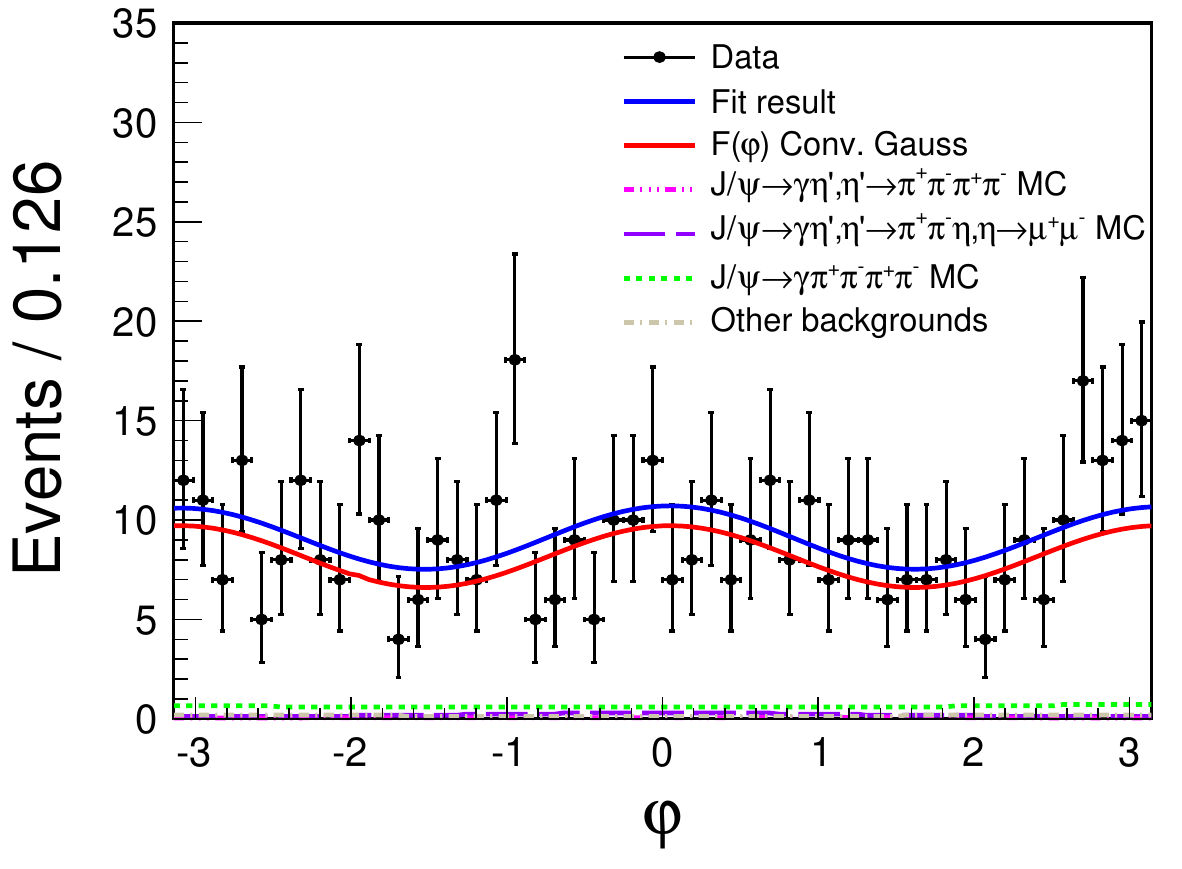}
	\put(25,60){{(b)}}
	\end{overpic}
	\caption{Fits to the $\varphi$ angle distribution for the (a) $\eta'\rightarrow\pi^+\pi^-e^+e^-$ and (b) $\eta'\rightarrow\pi^+\pi^-\mu^+\mu^-$ decays.
    The dots with error bars represent data, the blue solid line are the total fit results, and the red dashed histograms represents the function shapes of $F(\varphi)$ convolved with a Gaussian function. 
    For $\eta'\rightarrow\pi^+\pi^-e^+e^-$, the green area is the $J/\psi\rightarrow\gamma\eta', \eta'\rightarrow\gamma\pi^+\pi^-$ MC shape, and the brown dotted histogram is the background obtained from $\eta'$ sideband.
    For $\eta'\rightarrow\pi^+\pi^-\mu^+\mu^-$, the pink dotted histogram is the $J/\psi\rightarrow\gamma\eta', \eta'\rightarrow\pi^+\pi^-\pi^+\pi^-$ MC shape, the purple dashed histogram is the $J/\psi\rightarrow\gamma\eta', \eta'\rightarrow\pi^+\pi^-\eta, \eta\rightarrow\mu^+\mu^-$ MC shape, the green dashed histogram is the $J/\psi\rightarrow\gamma\pi^+\pi^-\pi^+\pi^-$ MC shape, and the golden dotted histogram is the sum of the other background MC shapes.}	
	\label{acp_fit}
\end{figure}

The mixed term between the $CP$-violating electric and $CP$-preserving magnetic form factors in the squared amplitude has a $\varphi$ dependence~\cite{KL_pipiee_1, BESIII_2020_etap_pipiee}. 
According to the squared invariant decay amplitude~\cite{1010_2378}, a fitting function is constructed as
\begin{equation}
F(\varphi) = 1+b\cdot \sin^{2}{\varphi}+c\cdot \sin{2\varphi},
\end{equation}
where $1 + b \cdot \sin^{2}{\varphi}$ is a dominant contribution from the magnetic term, and $c \cdot \sin{2\varphi}$ is from the mixing term, which is the $CP$-violating term.
The asymmetry parameter $\mathcal{A}_{CP}$ is defined as
\begin{equation}
\begin{aligned}
\mathcal{A}_{CP} &=\frac{1}{\Gamma}\int^{2\pi}_{0}\frac{\mathrm{d}\Gamma}{\mathrm{d}\varphi}\rm{sign}(\sin{2\varphi})\mathrm{d}\varphi\\
&= \frac{\int^{\pi}_{-\pi}F(\varphi)\rm{sign}(\sin{2\varphi})\mathrm{d}\varphi}{\int^{\pi}_{-\pi}F(\varphi)\mathrm{d}\varphi}\\
&= \frac{4c}{(2+b)\pi},
\label{Aphi}
\end{aligned}
\end{equation}
where $\mathrm{sign}(x)$ is a sign function. 
The parameters $b$ and $c$ and the asymmetry parameter $\mathcal{A}_{CP}$ are obtained by fitting the $\varphi$ angle distribution, as shown in Fig.~\ref{acp_fit}.
In the fit, the efficiency corrected $F(\varphi)$ function is convolved with a Gaussian function, $f(x;0,\sigma)=\frac{1}{\sigma\sqrt{2\pi}}\mathrm{exp}(-\frac{x^2}{2\sigma^2})$, to account for the $\varphi$ resolution, where $\sigma$ are determined to be 0.035 and 0.030 for $\eta'\rightarrow\pi^{+}\pi^{-}e^{+}e^{-}$ and $\eta'\rightarrow\pi^{+}\pi^{-}\mu^{+}\mu^{-}$, respectively, according to MC simulation.

The background treatment is the same as that in the TFF measurement in Section~\ref{sec:TFF mathod}, and the normalized background events are listed in Table~\ref{backgrounds}.
The final results are listed in Table~\ref{Acp_sum}, where the uncertainties are statistical only. These imply that no CP-violation evidence is found at the present level of statistics.

\begin{table}[htbp]
	\normalsize
	\renewcommand\arraystretch{1.1}
	\begin{center}
		\begin{tabular}{c|c|c|c|c}
        \hline
        \hline
        Decay channels & $b$ & $c$ &$\mathcal{A}_{CP} (\%)$  &$\chi^{2}/ndof$\\
        \hline
        $\eta'\rightarrow\pi^+\pi^-e^+e^-$ & $-0.72\pm0.01$ & $-0.002\pm0.007$ & $-0.21\pm0.73$ &82.8/96.0\\
        $\eta'\rightarrow\pi^+\pi^-\mu^+\mu^-$ & $-0.35\pm0.10$ & $0.008\pm0.061$ &$0.62\pm4.71$ &71.3/97.0\\
        \hline
        \hline
		\end{tabular}
	\end{center}
	\caption{The final results of $\mathcal{A}_{CP}$ for the $\eta'\rightarrow\pi^+\pi^-e^+e^-$ decay.}
	\label{Acp_sum}
\end{table}	

\section{SYSTEMATIC UNCERTAINTIES OF $\mathcal{B}$, $\mathcal{A}_{CP}$ AND TFF}
We consider possible sources of systematic uncertainties for the measurement of $\mathcal{B}$, $\mathcal{A}_{CP}$ and TFF. The corresponding contributions are discussed in detail below and listed in Table~\ref{sumsys}.

\begin{itemize}
\item Number of $J/\psi$ events:\\
The number of $J/\psi$ events is determined to be $(10087\pm44)\times10^6$ from the inclusive hadron events, and the uncertainty of this number is 0.4\%~\cite{BESIII_2021_njpsi}.
\item Branching fraction of $J/\psi\rightarrow\gamma\eta'$:\\ 
The uncertainty from the branching fraction of $J/\psi\rightarrow\gamma\eta'$ from the PDG~\cite{PDG_2022} is 1.3\%.
\item MDC tracking:\\ 
The data-MC efficiency difference for pion track-finding has been studied using a control sample of $J/\psi\rightarrow\pi^+\pi^-\pi^0$.
Since the muon mass is similar to that of the pion, we take the same efficiency difference as the pion for the tracking of the muon.
For the electron tracking, a mixed sample of $e^+e^-\rightarrow e^+e^-\gamma$ at the $J/\psi$ meson mass and $J/\psi\rightarrow e^+e^-\gamma_{\mathrm{FSR}}$ processes is used. 
The data-MC difference, $\Delta_{\rm{syst.}}$, is determined as a function of the particle momentum and the cosine of the polar angle.
We calculated the uncertainty of $\Delta_{\rm{syst.}}$ using the uncertainty propagation formula, and the result is 1.8\% as the systematic uncertainty for the $\mathcal{B}$ measurement.
Subsequently, each event in the MC samples is re-weighted by a factor $(1+\Delta_{\rm{syst.}})$. 
The $\mathcal{A}_{\mathrm{CP}}$ are recalculated with efficiencies determined from the re-weighted MC sample.
For the TFF measurement, a re-weighted MC sample is used to calculate the MC integral, and a group of new fit results are obtained by using the fit method as in section~\ref{sec:TFF mathod}.
The differences with the nominal results are taken as the systematic uncertainties.
\item PID:\\
The difference in the PID efficiencies for the $\eta'\rightarrow\pi^+\pi^-\mu^+\mu^-$ decay between data and MC simulation is evaluated in the same way as for the MDC tracking efficiency above. The control sample of $J/\psi\rightarrow\pi^+\pi^-\pi^0$ is used to study the pion PID.
For the muon PID, we take the same difference as that of the pion.
\item Photon detection efficiency: \\
The systematic uncertainty from the reconstruction of photons has been studied extensively in the process $e^+e^-\rightarrow\gamma\mu^+\mu^-$. 
The systematic uncertainty due to photon reconstruction efficiency, defined as the relative difference in efficiencies between data and MC simulation, is observed to be up to the level of 0.5\% in both the barrel and end-cap regions.
\item Kinematic fit: \\
To investigate the systematic uncertainty associated with the 4C kinematic fit, the track helix parameter correction method~\cite{BESIII_2013_kmfitfit} is used for $\eta'\rightarrow\pi^+\pi^-\mu^+\mu^-$. 
Half of the difference in the detection efficiencies with and without the helix correction is taken as the systematic uncertainty.
\item Combined PID and kinematic fit: \\
The control sample of $J/\psi\rightarrow\pi^+\pi^-\pi^0,\pi^0\rightarrow\gamma e^+e^-$ is used to study the systematic uncertainty due to the requirement of $\chi^2_{\rm{4C+PID}}<60$ for $\eta'\rightarrow\pi^+\pi^-e^+e^-$.
The difference of efficiencies with and without the $\chi^2_{\rm{4C+PID}}$ selection between data and MC simulation, 2.0\%, is taken as the systematic uncertainty.
\item Photon conversion veto: \\
The systematic uncertainty from photon conversion veto has been studied with a clean control sample of $J/\psi\rightarrow\pi^+\pi^-\pi^0,\pi^0\rightarrow\gamma e^+e^-$. The relative difference of efficiencies associated with the photon conversion rejection between data and MC simulation, 1.0\%, is taken as the systematic uncertainty~\cite{BESIII_2015_etap_gee}.
\item Generator model: \\
The signal MC sample was generated with $c_1-c_2=c_3=1$.
To estimate the systematic uncertainty from generator model, we used the fit results of the modified VMD model in Sec.~\ref{sec:TFF results} to produce the MC sample, and the difference of detection efficiency between the modified model ($c_1-c_2\neq c_3$) and the nominal models ($c_1-c_2=c_3=1$) is taken as the uncertainty.
\item $\eta'$ signal range: \\
To estimate the uncertainties from the $\eta'$ mass range, we perform alternative fits changing the lower and upper boundaries of the $\eta'$ mass range independently by 0.02 $\mathrm{GeV}/c^{2}$ for $\eta'\rightarrow\pi^+\pi^-e^+e^-$, 0.01 $\mathrm{GeV}/c^{2}$ for $\eta'\rightarrow\pi^+\pi^-\mu^+\mu^-$. 
The resultant largest differences to the nominal results are taken as the systematic uncertainties.
\item Background model: \\
In the fit, the events for some backgrounds are fixed according to the branching fractions from the PDG~\cite{PDG_2022}. 
To estimate the effect of the uncertainties of the used branching fractions, a set of random numbers have been generated within the uncertainty of each branching fraction. Using these random scaling parameters, a series of fits to the invariant mass distributions of $\pi^{+}\pi^{-}e^{+}e^{-}$ or $\pi^{+}\pi^{-}\mu^{+}\mu^{-}$ mass are performed. 
The largest changes of the signal yields are taken as the systematic uncertainties.
\item Signal shape: \\
In the $\mathcal{B}$ measurement, we use the double or three Gaussian functions instead of the signal MC shapes to fit the $\pi^{+}\pi^{-}l^{+}l^{-}$ mass spectrum.
The differences in the $\mathcal{B}$ values are taken as the systematic uncertainties.
\item Resolution: \\
In the $\mathcal{A}_{\mathrm{CP}}$ measurement, to estimate the uncertainty from the resolution, we perform alternative fits changing the resolution from $-1.0\sigma$ to $1.0\sigma$ with an interval of $0.25\sigma$ for the $\eta'\rightarrow\pi^{+}\pi^{-}e^{+}e^{-}$ and $\eta'\rightarrow\pi^{+}\pi^{-}\mu^{+}\mu^{-}$ decays.
The largest differences from the nominal results are taken as the systematic uncertainties of the asymmetry parameters.
\item Width of $\rho$ and $\omega$: \\
In the TFF measurement, the widths of $\rho$ and $\omega$ are taken as constants.
The differences of the fit results between fixed and free widths are taken as the systematic uncertainties. 
\end{itemize}

\begin{table}[htbp]
	\small
	\renewcommand\arraystretch{1.1}
	\begin{center}
		\begin{tabular}{c|c|c|c|c|c|c}
			\hline
			\hline
        \multirow{2}*{Source} & \multicolumn{3}{c|}{$\eta'\rightarrow\pi^+\pi^-e^+e^- (\%)$} & \multicolumn{3}{c}{$\eta'\rightarrow\pi^+\pi^-\mu^+\mu^- (\%)$} \\
        \cline{2-7} ~ & $\mathcal{B}$ & $\mathcal{A}_{CP}$ & $m_{V}$ & $\mathcal{B}$  & $\mathcal{A}_{CP}$ & $m_{V}$ \\ 
        \hline
        Number of $J/\psi$ events                     & $0.4$ & $-$     & $-$    & $0.4$ & $-$    & $-$\\
        $\mathcal{B}(J/\psi\rightarrow\gamma\eta')$   & $1.3$ & $-$     & $-$    & $1.3$ & $-$    & $-$\\
        MDC tracking                                  & $1.8$ & $-$     & $0.2$ & $0.7$ & $-$    & $2.0$\\
        PID                                           & $-$    & $-$     & $-$    & $1.3$ & $-$    & $2.0$\\
        Photon detection                              & $0.5$ & $-$     & $-$    & $0.5$ & $-$    & $-$\\
        4C kinematic fit                                 & $-$    & $3.7$  & $-$    & $0.6$ & $0.0$    & $1.1$\\
        Combined PID and kinematic fit                 & $2.0$ & $-$     & $2.0$ & $-$    & $-$    & $-$\\
        Photon conversion veto                        & $1.0$ & $-$     & $1.0$ & $-$    & $-$    & $-$\\
        Generator model                               & $0.1$ & $-$     & $-$    & $0.1$ & $-$    & $-$\\
        $\eta'$ mass range                            & $0.4$ & $-$& $-$ & $1.4$ & $-$& $-$\\
        Background model                              & $0.1$ & $0.1$  & $1.5$ & $0.5$ & $0.2$ & $3.5$\\
        Signal shape                              & $1.0$ & $-$  & $-$ & $1.4$ & $-$ & $-$\\
        Resolution                                    & $-$    & $4.0$  & $-$ & $-$       & $12.7$    & $-$\\
        Widths of $\rho$ and $\omega$     & $-$    & $-$     & $2.7$ & $-$    & $-$    & $3.0$\\
        \hline
        Total                                         & $3.4$ & $5.4$ & $3.8$ & $3.0$ & $12.7$ & $5.5$\\
			\hline
			\hline
		\end{tabular}
	\end{center}
	\caption{The systematic uncertainties in the measurements of $\mathcal{B}$, $\mathcal{A}_{CP}$ and $m_{V}$.}
	\label{sumsys}
\end{table}	

\section{UPPER LIMIT FOR ALPS IN DECAY OF $\eta'\rightarrow\pi^{+}\pi^{-}a,a\rightarrow e^{+}e^{-}$}
To study possible ALPs decaying into $e^{+}e^{-}$, we perform 50 simulations of the signal process $a\rightarrow e^{+}e^{-}$ with varied masses of $a$ in steps of $10$ MeV/$c^{2}$ from 0 to 500 MeV/$c^{2}$. 
Using the hypothesis of a pseudoscalar ALP, the signal process $\eta'\rightarrow\pi^{+}\pi^{-}a$ and $a\rightarrow e^{+}e^{-}$ is generated with the events distributed evenly in the phase space.
The MC shapes are determined from the sum of all background contributions plus the 50 groups of different signal MC shapes with $a$ masses from $0-500$ MeV/$c^{2}$.
The $\eta'$ signal region is defined as $0.945\ \mathrm{GeV}/c^{2} < M(\pi^{+}\pi^{-}e^{+}e^{-}) < 0.97\ \mathrm{GeV}/c^{2}$.

\begin{figure}[htbp]
	\centering
	\begin{overpic}[width=0.41\textwidth]{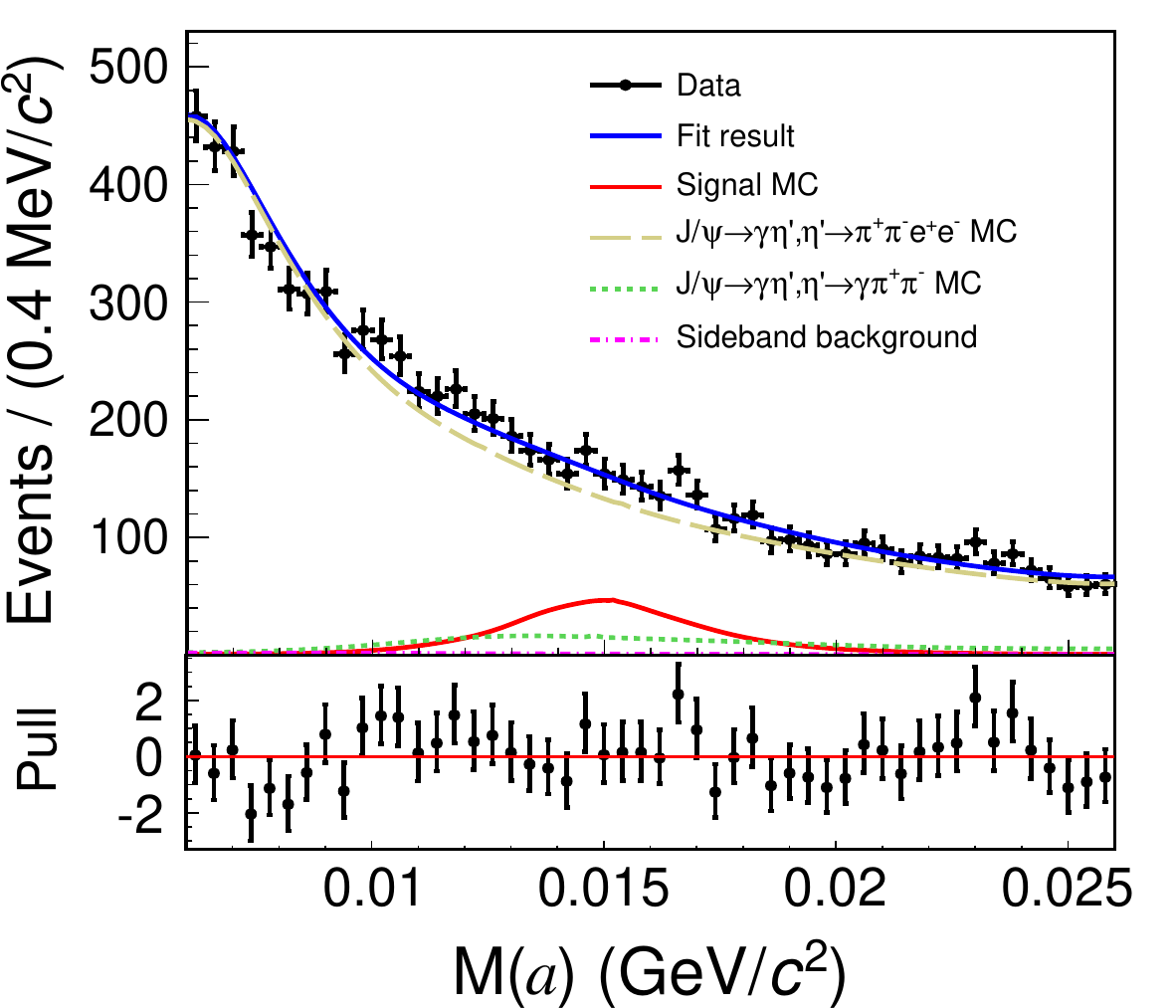}
	\put(80,75){{(a)}}
	\end{overpic}
	\begin{overpic}[width=0.48\textwidth]{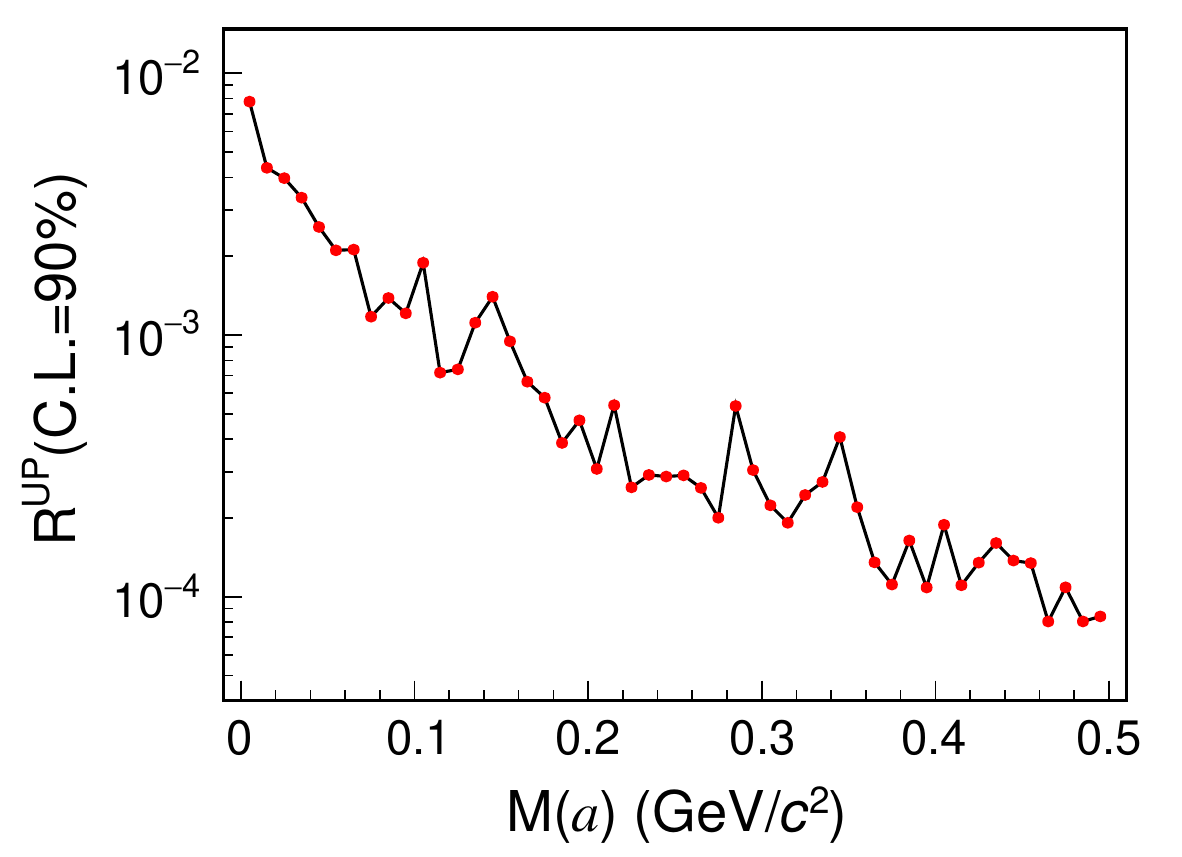}
	\put(80,60){{(b)}}
	\end{overpic}
	\caption{(a) Fit to the invariant mass distribution of M($a$) when the mass of $a$ is $15\ \mathrm{MeV}/c^{2}$. The dots with error bars represent data, and the blue solid line is the total fit result. The red dashed histogram represents the arbitrary normalized MC signal shape, which is scaled by a factor of 10. The golden dashed histogram is the $J/\psi\rightarrow\gamma\eta', \eta'\rightarrow\pi^{+}\pi^{-}e^{+}e^{-}$ MC shape. The green dashed histogram is the $J/\psi\rightarrow\gamma\eta', \eta'\rightarrow\gamma\pi^{+}\pi^{-}$ MC shape, and the pink dotted histogram is the background obtained from the $\eta'$ sideband.
	(b) Upper limit on the relative branching ratio at the 90\% C.L. for different $a$ masses.}	
	\label{upper_smear}
\end{figure}

We consider possible sources for multiplicative systematic uncertainties of the upper limit for ALPs, such as number of $J/\psi$ events (0.4\%), branching fraction of $J/\psi\rightarrow\gamma\eta'$ (1.3\%), MC statistics (0.1\%), MDC tracking (1.3\%-7.5\%), photon detection efficiency (0.5\%), combined PID and kinematic fit (2.0\%) and photon conversion veto (1.0\%). 
The total multiplicative systematic uncertainties are $3.0\%-7.9\%$.
The additive systematic uncertainties are considered by alternative fit ranges and background models. The maximum number of signal events among the different fit scenarios is adopted to calculate the upper limit of the signal yield.
Since no evident $a$ signal seen in the $M(e^+e^-)$ distribution, we compute upper limits on the branching ratio, $R^{\mathrm{UP}}=\frac{\mathcal{B}(\eta'\rightarrow\pi^{+}\pi^{-}a)\cdot\mathcal{B}(a\rightarrow e^{+}e^{-})}{\mathcal{B}(\eta'\rightarrow\pi^{+}\pi^{-}e^{+}e^{-})}$, at the 90\% Confidence Level (C.L.) as a function of $M(a)$.
The upper limits on the number of $a$ signal events at the 90\% C.L. are obtained according to the Bayesian method~\cite{Zhu:2008ca} by smearing the likelihood curve using a Gaussian function with a width of the systematic uncertainty as
\begin{equation}
L'(N) = \int^{1}_{0} L\left(\frac{S}{\hat{S}}N\right)\mathrm{exp}\left[-\frac{(S-\hat{S})^{2}}{2\sigma_{S}^{2}}\right]dS,
\end{equation}
where $L$ and $L'(N)$ are the likelihood curves before and after taking into account the systematic uncertainties; $\hat{S}$ is the nominal efficiency and $\sigma_{S}$ is its systematic uncertainty. 
As shown in Fig.~\ref{upper_smear}, the limits on the branching ratio are established at the level of $(0.1-7.8)\times10^{-3}$, and the significance for each case is less than $0.5 \sigma$.

\section{SUMMARY}
\begin{figure}[htbp]
	\centering
	\begin{overpic}[width=0.49\textwidth]{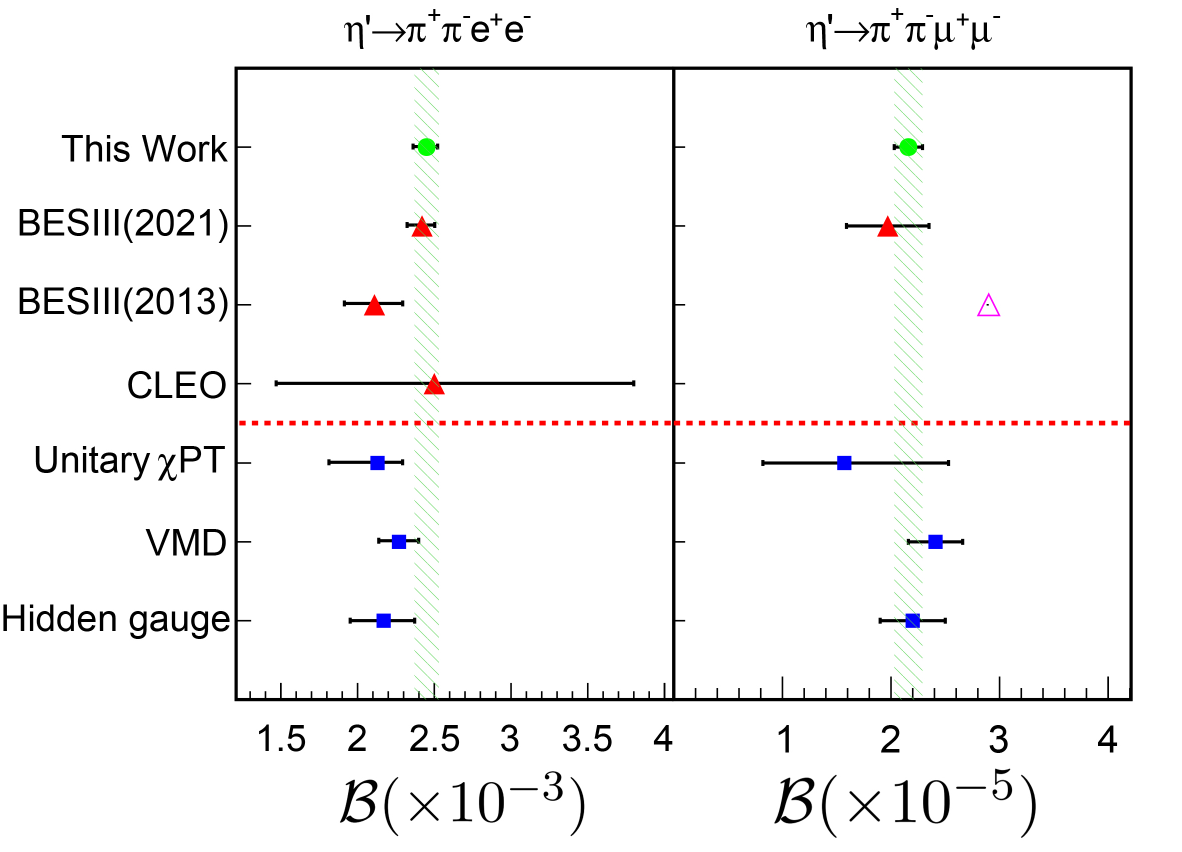}
	\put(85,60){{(a)}}
	\end{overpic}
	\begin{overpic}[width=0.49\textwidth]{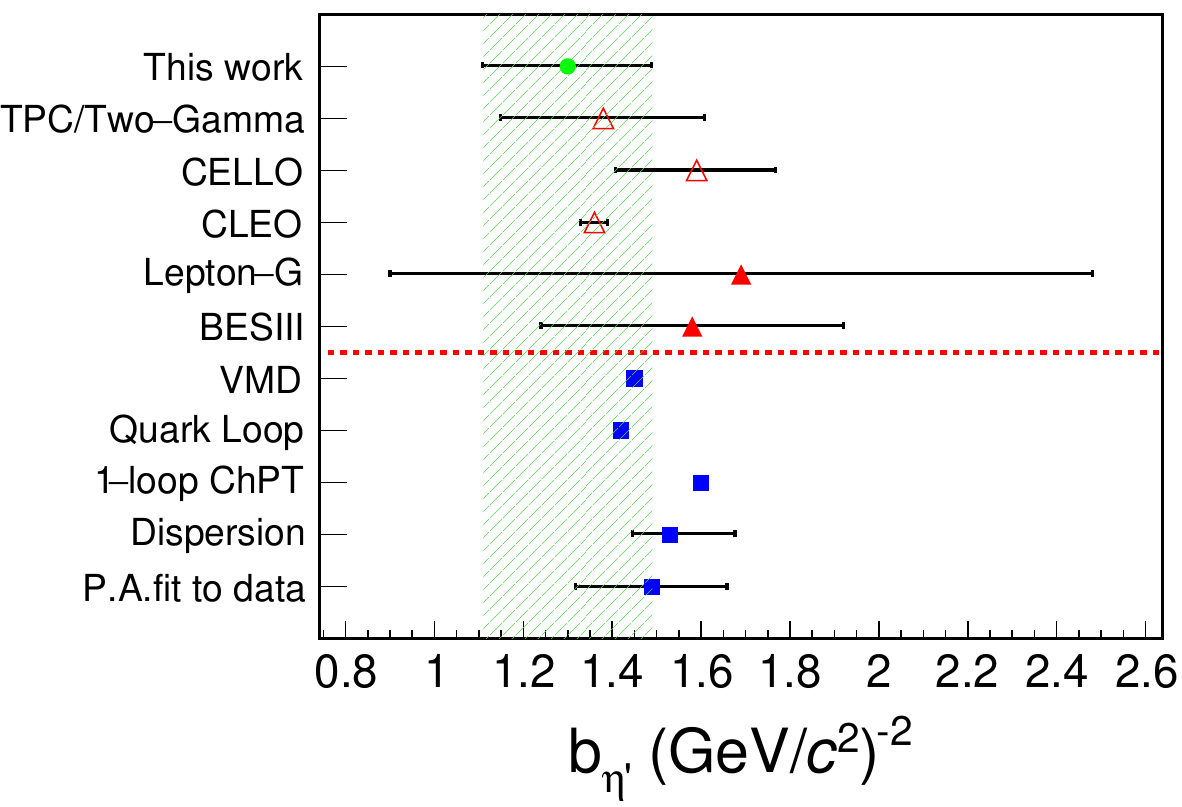}
	\put(85,60){{(b)}}
	\end{overpic}
	\caption{(a) The branching fraction of the $\eta'\rightarrow\pi^{+}\pi^{-}l^{+}l^{-}$ from different theoretical calculations~\cite{1010_2378, chiral_unitary_approach} (blue squares), the other experiments~\cite{BESIII_2013_etap_pipill, BESIII_2020_etap_pipimumu, BESIII_2020_etap_pipiee, CLEO_2008} (red solid and pink (90\% C.L.) hollow triangles), and the results in our work (green dots). The upper limit of $\mathcal{B}(\eta'\rightarrow\pi^{+}\pi^{-}\mu^{+}\mu^{-})<2.4\times10^{-4}$ at the 90\% C.L. is provided by the CLEO experiment~\cite{CLEO_2008}.
	(b) Slope parameter $b_{\eta'}$ for $\eta'$ TFF from different theoretical calculations~\cite{TFF_VMD_1981, TFF_quark_1983, TFF_theory_1992, TFF_dispersive_2013, TFF_theory_2014}(blue squares), time-like measurements~\cite{TFF_LeptonG_1980_1, TFF_LeptonG_1980_2, Event_generators_2015_etap_gmumu}(red solid triangles), space-like measurements~\cite{TFF_TPCTwoGamma_1990, TFF_CELLO_1990, TFF_CLEO_1997} (red hollow triangles), and the results in our work (green dots).}	
	\label{sumcom}
\end{figure}

With a sample of $(10087\pm44)\times10^{6}$ $J/\psi$ events, the branching fractions of $\eta'\rightarrow\pi^{+}\pi^{-}l^{+}l^{-}$ (with $l=e,\mu$) are determined to be $\mathcal{B}(\eta'\rightarrow\pi^{+}\pi^{-}e^{+}e^{-})=(2.45\pm0.02(\rm{stat.})\pm0.08(\rm{syst.}))\times10^{-3}$ and $\mathcal{B}(\eta'\rightarrow\pi^{+}\pi^{-}\mu^{+}\mu^{-})=(2.16\pm0.12(\rm{stat.})\pm0.06(\rm{syst.}))\times10^{-5}$, which are in good agreement with the theoretical predictions~\cite{1010_2378, chiral_unitary_approach} and the previous measurements~\cite{BESIII_2013_etap_pipill, BESIII_2020_etap_pipimumu, BESIII_2020_etap_pipiee, CLEO_2008}, as shown in Fig.~\ref{sumcom} (a).
The ratio of the branching fractions of these two decays is calculated to be $\frac{\mathcal{B}(\eta'\rightarrow\pi^{+}\pi^{-}e^{+}e^{-})}{\mathcal{B}(\eta'\rightarrow\pi^{+}\pi^{-}\mu^{+}\mu^{-})} = 113.4\pm0.9(\rm{stat.})\pm3.7(\rm{syst.})$.

Additionally, we measure the TFF parameters of the $\eta'$ decays, which are determined from the invariant decay amplitude of the reaction $\eta'\rightarrow\pi^+\pi^-l^+l^-$. 
Our results are summarized in Table~\ref{TFF_pipiee}. 
Since the previous experiment measured the parameters value of $c_1 - c_2$ and $c_3$ are approximately equal to $1$~\cite{ci_2010}, we compared the results of Model I with other theoretical calculations and experimental measurements, which shown in Fig.~\ref{sumcom} (b).
In addition, we calculate a weighted average of $b_{\eta'}=1.30\pm0.19\ (\mathrm{GeV}/c^{2})^{-2}$ for $\eta'\rightarrow\pi^{+}\pi^{-}e^{+}e^{-}$ and $\eta'\rightarrow\pi^{+}\pi^{-}\mu^{+}\mu^{-}$ combined, where the uncertainty is obtained by combining statistical and systematic uncertainties.
The value of $b_{\eta'}$ measured in this work is consistent with the VMD theoretical calculations~\cite{TFF_VMD_1981} and the previous BESIII result~\cite{Event_generators_2015_etap_gmumu}, as shown in Fig.~\ref{sumcom} (b).
It indicates that the theoretical model used (Model I) is able to reasonably describe the process.

The $CP$-violating asymmetries of $\eta'\rightarrow\pi^{+}\pi^{-}l^{+}l^{-}$ are determined to be $\mathcal{A}_{CP}(\eta'\rightarrow\pi^{+}\pi^{-}e^{+}e^{-})=(-0.21\pm0.73(\rm{stat.})\pm0.01(\rm{syst.}))\%$ and $\mathcal{A}_{CP}(\eta'\rightarrow\pi^{+}\pi^{-}\mu^{+}\mu^{-})=(0.62\pm4.71(\rm{stat.})\pm0.08(\rm{syst.}))\%$. These imply that no $CP$-violation evidence is found at the present level of statistics.

Finally, we perform a search for an ALP in the $e^{+}e^{-}$ invariant mass spectrum, no significant signal is observed and the 90\% C.L. upper limits of $\mathcal{B}^{\mathrm{UP}}$ are shown in Fig.~\ref{upper_smear}.

\acknowledgments
The BESIII Collaboration thanks the staff of BEPCII and the IHEP computing center for their strong support. This work is supported in part by National Key R\&D Program of China under Contracts Nos. 2020YFA0406300, 2020YFA0406400; National Natural Science Foundation of China (NSFC) under Contracts Nos. 11635010, 11735014, 11835012, 11935015, 11935016, 11935018, 11961141012, 12025502, 12035009, 12035013, 12061131003, 12192260, 12192261, 12192262, 12192263, 12192264, 12192265, 12221005, 12225509, 12235017; the Chinese Academy of Sciences (CAS) Large-Scale Scientific Facility Program; the CAS Center for Excellence in Particle Physics (CCEPP); Joint Large-Scale Scientific Facility Funds of the NSFC and CAS under Contract No. U1832207; CAS Key Research Program of Frontier Sciences under Contracts Nos. QYZDJ-SSW-SLH003, QYZDJ-SSW-SLH040; 100 Talents Program of CAS; The Institute of Nuclear and Particle Physics (INPAC) and Shanghai Key Laboratory for Particle Physics and Cosmology; European Union's Horizon 2020 research and innovation programme under Marie Sklodowska-Curie grant agreement under Contract No. 894790; German Research Foundation DFG under Contracts Nos. 455635585, Collaborative Research Center CRC 1044, FOR5327, GRK 2149; Istituto Nazionale di Fisica Nucleare, Italy; Ministry of Development of Turkey under Contract No. DPT2006K-120470; National Research Foundation of Korea under Contract No. NRF-2022R1A2C1092335; National Science and Technology fund of Mongolia; National Science Research and Innovation Fund (NSRF) via the Program Management Unit for Human Resources \& Institutional Development, Research and Innovation of Thailand under Contract No. B16F640076; Polish National Science Centre under Contract No. 2019/35/O/ST2/02907; The Swedish Research Council; U. S. Department of Energy under Contract No. DE-FG02-05ER41374.

\bibliography{apssamp}

\newpage

\noindent$\Large\textbf{The BESIII Collaboration}$ \\
\\
\begin{small}
M.~Ablikim$^{1}$, M.~N.~Achasov$^{4,b}$, P.~Adlarson$^{75}$, O.~Afedulidis$^{3}$, X.~C.~Ai$^{80}$, R.~Aliberti$^{35}$, A.~Amoroso$^{74A,74C}$, Q.~An$^{71,58}$, Y.~Bai$^{57}$, O.~Bakina$^{36}$
% ...
\\
\vspace{0.2cm}
(BESIII Collaboration)\\
\vspace{0.2cm} 
{\it
$^{1}$ Institute of High Energy Physics, Beijing 100049, People's Republic of China\\
% ...
$^{3}$ Bochum  Ruhr-University, D-44780 Bochum, Germany\\
$^{4}$ Budker Institute of Nuclear Physics SB RAS (BINP), Novosibirsk 630090, Russia\\
% ...
$^{35}$ Johannes Gutenberg University of Mainz, Johann-Joachim-Becher-Weg 45, D-55099 Mainz, Germany\\
$^{36}$ Joint Institute for Nuclear Research, 141980 Dubna, Moscow region, Russia\\
% ...
$^{57}$ Southeast University, Nanjing 211100, People's Republic of China\\
$^{58}$ State Key Laboratory of Particle Detection and Electronics, Beijing 100049, Hefei 230026, People's Republic of China\\
% ...
$^{71}$ University of Science and Technology of China, Hefei 230026, People's Republic of China\\
% ...
$^{74}$ University of Turin and INFN, (A)University of Turin, I-10125, Turin, Italy; (B)University of Eastern Piedmont, I-15121, Alessandria, Italy; (C)INFN, I-10125, Turin, Italy\\
$^{75}$ Uppsala University, Box 516, SE-75120 Uppsala, Sweden\\
% ...
$^{80}$ Zhengzhou University, Zhengzhou 450001, People's Republic of China\\

\vspace{0.2cm}
% ...
$^{b}$ Also at the Novosibirsk State University, Novosibirsk, 630090, Russia\\
% ...

}
\end{small}
\newpage

\end{document}